\pdfoutput=1
\newcommand*{\ATLASLATEXPATH}{}
\documentclass[cernpreprint, atlasdraft=false, texlive=2018, txfonts, UKenglish, LANGEDIT=true, LANGSHOW=false]{\ATLASLATEXPATH atlasdoc}

\usepackage[subfigure=true]{\ATLASLATEXPATH atlaspackage}
\usepackage[backref=false]{\ATLASLATEXPATH atlasbiblatex}
\usepackage{longtable}
 
\usepackage{\ATLASLATEXPATH atlasphysics}
 
\defbibentryset{IBL}{Capeans:1291633,CERN-LHCC-2012-009}
 
\graphicspath{{logos/}{figures/}}
 
\usepackage{ANA-SUSY-2018-23-PAPER-defs}

\AtlasTitle{Search for direct production of electroweakinos in final states with missing transverse momentum and a Higgs boson decaying into photons in $pp$ collisions at $\sqrt{s}=13$~TeV with the ATLAS detector}
 
\AtlasAbstract{
A search for a chargino--neutralino pair decaying via the 125~GeV Higgs boson into photons is presented.
The study is based on the data collected between 2015 and 2018 with the ATLAS detector at the LHC, corresponding to an integrated luminosity of $139~\mathrm{fb}^{-1}$ of $pp$ collisions at a centre-of-mass energy of 13~TeV.
No significant excess over the expected background is observed.
Upper limits at 95\% confidence level for a massless $\tilde{\chi}^{0}_{1}$ are set on several electroweakino production cross-sections
and the visible cross-section for beyond the Standard Model processes.
In the context of simplified supersymmetric models, 95\% confidence-level limits of up to 310~GeV in $m(\tilde{\chi}^{\pm}_{1}/\tilde{\chi}^{0}_{2})$, where $m(\tilde{\chi}^{0}_{1})=0.5$~GeV, are set.
Limits at 95\% confidence level are also set on the $\tilde{\chi}^{\pm}_{1}\tilde{\chi}^{0}_{2}$ cross-section in
the mass plane of $m(\tilde{\chi}^{\pm}_{1}/\tilde{\chi}^{0}_{2})$ and $m(\tilde{\chi}^{0}_{1})$, and on scenarios with gravitino as the lightest supersymmetric particle.
Upper limits at the 95\% confidence-level are set on the higgsino production cross-section.
Higgsino masses below 380~GeV are excluded for the case of the higgsino fully decaying into a Higgs boson and a gravitino.
}

\author{The ATLAS Collaboration}
 
\AtlasRefCode{SUSY-2018-23}
 
\AtlasJournalRef{JHEP 10 (2020) 005}
\AtlasDOI{10.1007/JHEP10(2020)005}

\PreprintIdNumber{CERN-EP-2019-204}

\AtlasJournal{JHEP}

\hypersetup{pdftitle={ATLAS document},pdfauthor={The ATLAS Collaboration}}
 
\begin{document}
 
\maketitle

\clearpage
\section{Introduction}
\label{sec:intro}
 
Theoretical and experimental arguments suggest that the Standard Model (SM) is an effective
theory valid up to a certain energy scale. The SM Higgs boson, denoted by $h$, is observed by the ATLAS and CMS collaborations~\cite{HIGG-2012-27,CMS-HIG-12-028,HIGG-2014-14,HIGG-2015-07}.
The Higgs boson mass is strongly sensitive to quantum corrections from physics at very high energy scales and demands a high level of fine-tuning, known as the hierarchy problem~\cite{Sakai:1981gr,Dimopoulos:1981yj,Ibanez:1981yh,Dimopoulos:1981zb}.
Supersymmetry (SUSY)~\cite{Golfand:1971iw,Volkov:1973ix,Wess:1974tw,Wess:1974jb,Ferrara:1974pu,Salam:1974ig}
resolves the hierarchy problem by introducing, for each known particle state, a new partner (superpartner) that shares the same mass and internal quantum numbers with the exception of spin if supersymmetry is unbroken.
However, these superpartners have not been observed, so SUSY must be a broken symmetry and the mass scale of the
supersymmetric particles is as yet undetermined.
The possibility of a supersymmetric dark matter candidate~\cite{Goldberg:1983nd,Ellis:1983ew} is
closely related to the conservation of \emph{R}-parity~\cite{Farrar:1978xj}. Under the \emph{R}-parity conservation
hypothesis, the lightest supersymmetric particle (LSP) is stable.
If the LSP is weakly interacting, it may provide a viable dark matter candidate.
The nature of the LSP is defined by the mechanism that spontaneously breaks supersymmetry and the parameters of
the chosen theoretical framework.
 
In the SUSY scenarios considered as a first benchmark in this paper, the LSP is the lightest of the neutralinos $\tilde{\chi}_{j}^{0}$ $(j=1,2,3,4)$ that,
together with the charginos $\tilde{\chi}_{i}^{\pm}$ $(i=1,2)$, represent the mass eigenstates formed from
the mixture of the $\gamma$, $\Wboson$, $\Zboson$ and Higgs bosons' superpartners (the winos, binos and higgsinos).
The neutralinos and charginos are collectively referred to as \textit{electroweakinos}.
Specifically, the electroweakino mass eigenstates are designated in order of increasing mass.
Naturalness considerations~\cite{Barbieri:1987fn,deCarlos:1993yy} suggest that the
lightest of the charginos and neutralinos have masses near the electroweak scale.
Their direct production may be the dominant mechanism at the Large Hadron Collider
(LHC) if the superpartners of the gluons and quarks are heavier than a few \TeV.
In SUSY models where the heaviest (pseudoscalar, charged) minimal supersymmetric Standard Model (MSSM) Higgs bosons and
the superpartners of the leptons have masses larger than those of the lightest chargino and next-to-lightest
neutralino, the former might decay into the \nn and a \Wboson boson ($\chargino \rightarrow W \nn$),
while the latter could decay into the \nn and the lightest MSSM Higgs boson or \Zboson boson
($\twoneutralino \rightarrow h / Z \nn$)~\cite{Fayet:1976et,Fayet:1977yc,Farrar:1978xj}.
The decay via the Higgs boson is dominant for many choices of parameters as long as the mass-splitting between the two
lightest neutralinos is larger than the Higgs boson mass and the higgsinos are heavier than the
winos. SUSY models of this kind
could provide a possible explanation for the discrepancy between measurements of the muon's anomalous magnetic moment $g-2$
and SM predictions~\cite{Das:2014kwa,Ajaib:2015yma,Chakraborti:2015mra,Endo:2017zrj}.
 
This paper presents a search in proton--proton ($pp$) collisions produced at the LHC at a centre-of-mass energy $\sqrt{s} = 13$~\TeV\ for the
direct pair production of electroweakinos that promptly decay into the LSP, producing at least one Higgs boson, decaying into two
photons in each event.
The primary model, for which the search is optimised, involves the production of a
chargino in association with a next-to-lightest neutralino, which promptly decay as $\chargino \rightarrow W \nn$ and $\twoneutralino \rightarrow h \nn$
respectively (see Figure~\ref{fig:N1N1_diagram}(a)),
the $\nn$ in the final state leading to a signature of missing transverse momentum, whose magnitude is denoted by $\met$.
A simplified SUSY model~\cite{Alwall:2008ag,Alves:2011wf} is considered for the optimisation of the search and the interpretation of results.
The $\chargino \rightarrow W \nn$ and $\twoneutralino \rightarrow h \nn$ decays are each assumed to have a 100\% branching fraction.
The Higgs boson branching fractions are assumed to be the same as in the SM~\cite{deFlorian:2016spz}.
The result from the CMS experiment using an integrated luminosity of 77.5 \ifb\ of \(pp\) collision data is given in Ref.~\cite{Sirunyan:2019iwo}.
Although the branching fraction of the SM Higgs boson decaying into a pair of photons is small,
the diphoton system presented in this paper falls in a narrower mass range around the Higgs boson mass than in Refs.~\cite{Aad:2019vvf,Sirunyan:2018ubx}
where the SM Higgs boson decay into a pair of $b$-quarks.
With the diphoton trigger, this channel is more sensitive in the low $\met$ region than the
channel with the SM Higgs boson decaying into a pair of $b$-quarks, which relies on the high $\met$ trigger.
In addition, a prior search from ATLAS~\cite{SUSY-2017-01} for this process making
use of 36.1 \ifb\ of \(pp\) collision data, based purely on leptonic
decays of the $W$ boson, observed a small excess of events above the SM prediction. This prior search is also updated to the full Run 2 data, and referred to as `follow-up' analysis.
 
\begin{figure}[htbp!]
\centering
\begin{tabular}{cccc}
\subfigure[]{\includegraphics[width=0.28\textwidth]{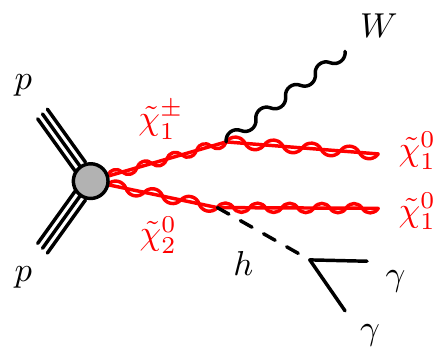}}
\subfigure[]{ \includegraphics[width=0.28\textwidth]{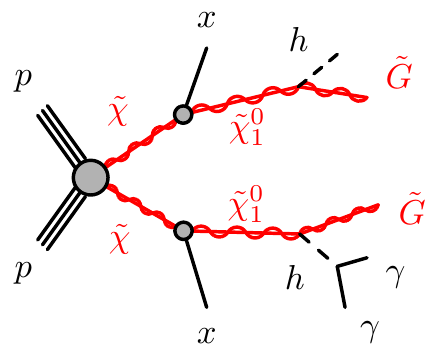}}
\end{tabular}
\caption{
Signal diagrams illustrating (a) $\cc\nnn$ production,
and (b) a higgsino production mode from a GMSB model: 
$\nn \to h \tilde{G}$.
For $\cc\nnn$ production, the lightest chargino (\chargino) and next-to-lightest
neutralino (\nnn) are nearly mass degenerate. In the higgsino models, the two lightest neutralinos, \nn and \nnn, and the lightest chargino \cc are approximately mass degenerate, and
the \nn is the lightest of the four nearly degenerate higgsino states, $x$ is the particle with low momentum from the promptly decay of \cc and \nnn.
}
\label{fig:N1N1_diagram}
\end{figure}
 
The analysis optimised for the search for \chargino\nnn production is also used to search for a
gauge-mediated supersymmetry breaking (GMSB)~\cite{Dine:1981gu,AlvarezGaume:1981wy,Nappi:1982hm} scenario featuring direct production of pairs of higgsinos~\cite{Khachatryan:2014mma,Matchev:1999ft,Ruderman:2011vv}, collectively denoted by $\tilde{\chi}\tilde{\chi}$.
In this model, the two lightest neutralinos, \nn and \nnn, and the lightest chargino \cc are approximately mass degenerate, and
the \nn is the lightest of the four nearly degenerate higgsino states. The masses are assumed to be related by $m(\cc) = m(\nnn) = m(\nn)+1~\GeV$.
The effective cross-section for higgsino production is a combination of the cross-sections for \nn\nnn, \nn\cc, \nnn\cc, and \cc\ccmp production.
In the GMSB scenarios considered (Figure~\ref{fig:N1N1_diagram}(b)),
a 100\% branching fraction for $\nn \to h \tilde{G}$ is assumed, where $\tilde{G}$ indicates the gravitino (the superpartner of the graviton).
This scenario is denoted by $h \tilde{G} h \tilde{G}$ in the following.
In this scenario, $\tilde{G}$ in the final state is stable, weakly interacting, and nearly massless, which leads to an $\met$ signature.
 
The general strategy of the analysis is to search for beyond the Standard Model (BSM) events by using a simultaneous signal-plus-background fit to the full $m_{\gamma\gamma}$ spectrum for different categories.
The paper is organised as follows.
Section~\ref{sec:detector} presents a brief description of the ATLAS detector.
Section~\ref{sec:datamc} introduces the data, the signal and background Monte Carlo (MC) simulation samples used.
Section~\ref{sec:objsel} outlines the event reconstruction, while
Section~\ref{sec:evtsel} explains the optimisation of the event selection and categorisation.
Section~\ref{sec:param} discusses the signal and background modelling.
Section~\ref{sec:sys} summarises the experimental and theoretical systematic uncertainties that affect the results.
Section~\ref{sec:results} describes the results and their interpretations, and
conclusions are drawn in Section~\ref{sec:conclusion}.
 
\section{ATLAS detector}
\label{sec:detector}
 
The ATLAS detector~\cite{PERF-2007-01} is a multipurpose particle detector with a forward--backward symmetric cylindrical
geometry and nearly 4$\pi$ coverage in solid angle.\footnote{ATLAS uses a right-handed coordinate system with its origin at the nominal
interaction point in the centre of the detector. The positive $x$-axis is defined by the direction from the interaction point to the centre of the LHC ring,
with the positive $y$-axis pointing upwards, while the beam direction defines the $z$-axis. Cylindrical coordinates
$(r,\phi)$ are used in the transverse plane, $\phi$ being the azimuthal angle around the $z$-axis. The pseudorapidity $\eta$ is
defined in terms of the polar angle $\theta$ by $\eta=-\ln\tan(\theta/2)$. Rapidity is defined as $y$ = $0.5 \ln[(E + p_z)/(E-p_z)]$ where $E$
denotes the energy and $p_z$ is the component of the momentum along the beam direction. The angular distance $\Delta R$ is defined as $\sqrt{(\Delta y)^2 + (\Delta \phi)^2}$.}
The inner tracking detector (ID) consists of pixel and microstrip silicon detectors
covering the pseudorapidity region $|\eta|<2.5$, surrounded by a transition radiation tracker
that enhances electron identification in the region $|\eta|<2.0$.
A new inner pixel layer, the insertable B-layer~\cite{IBL, Abbott:2018ikt}, was added at a mean radius of 3.3~cm during the period between Run~1 and Run~2 of the LHC\@.
The inner detector is surrounded by a thin superconducting solenoid providing an axial 2~T magnetic field and by a lead/liquid-argon electromagnetic (EM) sampling calorimeter covering $|\eta|<3.2$, with a fine-granularity region up to $|\eta|=2.5$. A steel/scintillator-tile hadronic sampling calorimeter provides coverage in the central pseudorapidity
range ($|\eta|<1.7$). The endcap and forward regions ($1.5<|\eta|<4.9$) of the hadronic calorimeter are made of liquid-argon active layers with either copper or tungsten
as the absorber material. A muon spectrometer with an air-core toroid magnet system surrounds the calorimeters.
Three layers of high-precision tracking chambers provide coverage in the range $|\eta|<2.7$, while dedicated fast chambers allow triggering
in the region $|\eta|<2.4$. The ATLAS trigger system consists of a hardware-based first-level trigger followed by a software-based high-level trigger~\cite{TRIG-2016-01}.
 
\section{Data and simulation samples}
\label{sec:datamc}

The analysis uses $pp$ collision data with a bunch crossing interval of $25$~ns,
collected from 2015 to 2018 at $\sqrt{s}= 13$~\TeV. Only events that were recorded
in stable beam conditions, when relevant detector components were functioning
properly, are considered. A diphoton trigger~\cite{Aad:2019wsl} was used to collect the events by
requiring two reconstructed photon candidates with transverse energies (\ET) of at least $35$~\GeV\ and $25$~\GeV\ for the $\et$-ordered leading and
subleading photons respectively. The trigger efficiency relative to the
offline-reconstructed photons 
was 99\%.
The data sample corresponds to an integrated luminosity of $139.0 \pm 2.4 ~\mathrm{fb}^{-1}$.
There are, on average, 25 to 38 interactions in the same bunch crossing (in-time pile-up)
in the data sample.
 
The MC simulation of signal and background processes is used  to optimise the selection criteria, estimate uncertainties and study the shapes
of the signal and background diphoton invariant mass (\mgg) distributions. Signal
events were generated with up to two additional partons in the matrix element using \amcnlo\_aMC@NLO 2.6.2~\cite{Alwall:2014hca} at leading order (LO) in quantum chromodynamics (QCD)
using the NNPDF3.0LO~\cite{Ball:2014uwa} parton distribution function (PDF) set and CKKW-L merging scheme.
Parton showering and hadronisation were handled by the \pythiaEight~\cite{Sjostrand:2007gs}
event generator with the A14~\cite{ATL-PHYS-PUB-2014-021} set of tuned
parameters (tune), using the NNPDF2.3LO PDF set~\cite{Ball:2012cx}.
MC samples for the $\cc\nnn$ production were generated assuming $m(\cc)=m(\nnn)$ for a range of values of $m(\nn)$.
As shown in Figure~\ref{fig:genplot}(a), the transverse momentum (\pt) distribution of the $\nn\nn$ system is broader for higher values of the difference $m(\cc/\nnn)-m(\nn)$.
The \pt distributions of the $\tilde{G}\tilde{G}$ system for the higgsino production of $h \tilde{G} h \tilde{G}$ are presented in Figure~\ref{fig:genplot}(b).
The MC samples include \nn\nnn, \nn\cc, \nnn\cc, and \cc\ccmp production.
The kinematic distributions depend strongly on the mass of the $\nn$, where the mass of the $\tilde{G}$ is assumed to be 1~\MeV.
 
Signal cross-sections were calculated to NLO in the strong coupling constant, $\alphas$, adding the resummation of soft gluon emission at next-to-leading-logarithm accuracy (NLO+NLL)~\cite{Debove:2010kf,Fuks:2012qx,Fuks:2013vua,Fiaschi:2018hgm,Beenakker:1999xh}.
The nominal cross-section and its uncertainty are taken from an envelope of cross-section predictions using different PDF sets and factorisation and renormalisation scales, as described in Ref.~\cite{Borschensky:2014cia}.
 
\begin{figure}[!htpb]
\centering
\begin{tabular}{cccc}
\subfigure[]{\includegraphics[width=0.45\textwidth]{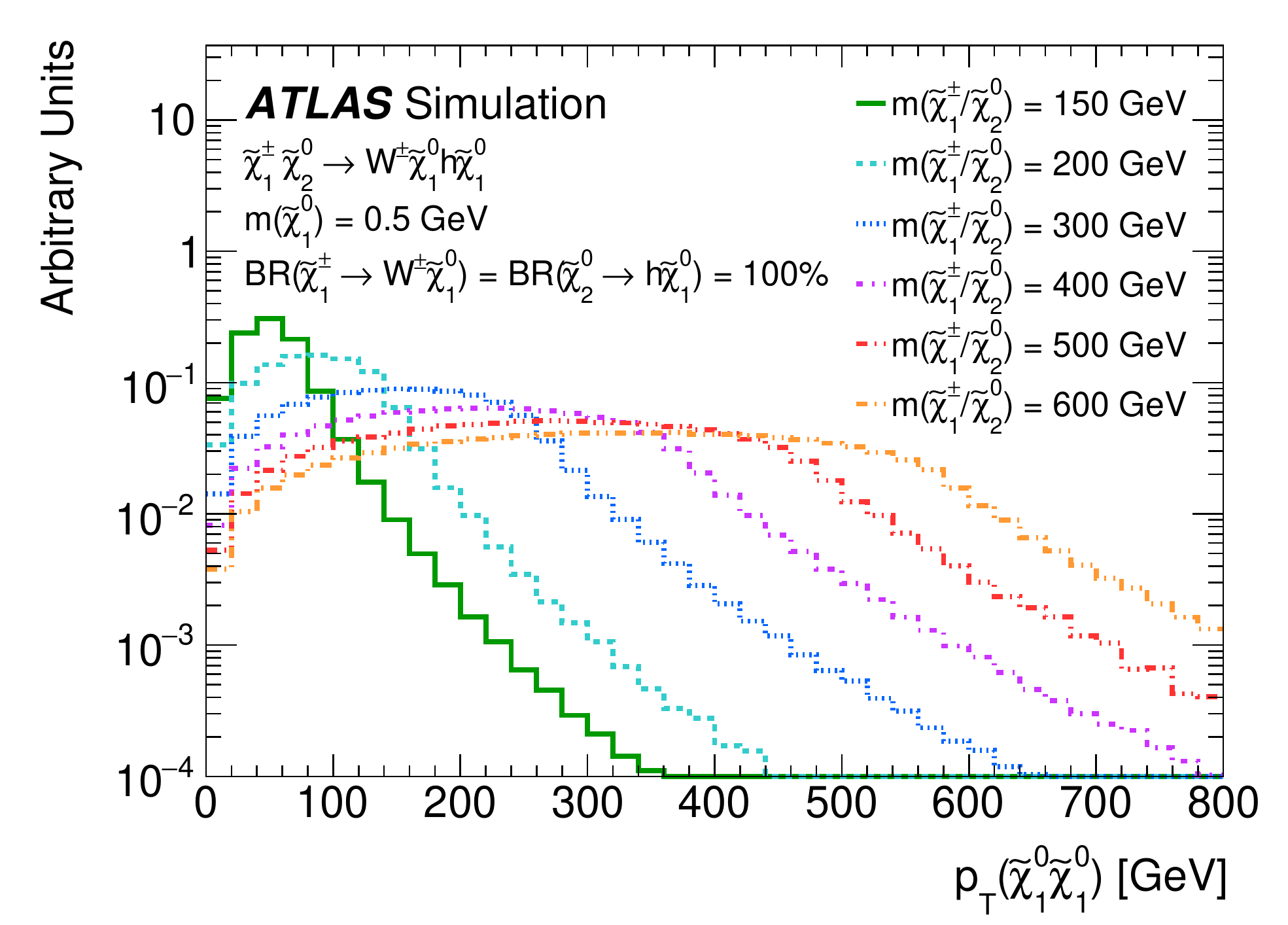}}
\subfigure[]{\includegraphics[width=0.45\textwidth]{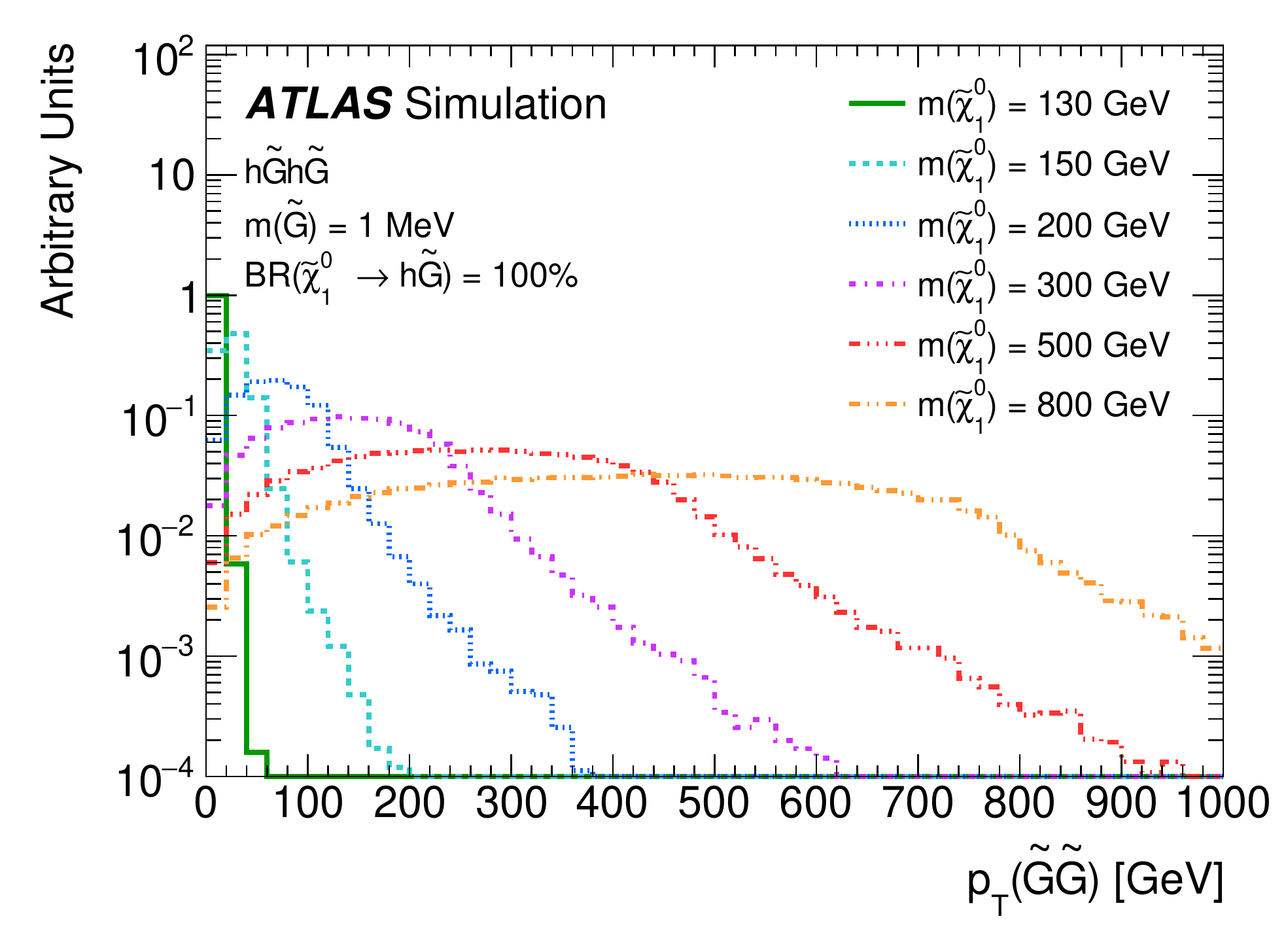}}
\end{tabular}
\caption{The \pt distribution of (a) the $\nn\nn$  in $W^{\pm}\nn h\nn$ production and (b) $\tilde{G}\tilde{G}$ in $h\tilde{G} h \tilde{G}$ production.
}
\label{fig:genplot}
\end{figure}
 
The dominant backgrounds are resonant SM $h \rightarrow \gamma\gamma$ processes, and non-resonant processes that include
$\gamma\gamma$, $\gamma$+jets, $V\gamma$ ($V$=$W$, $Z$) and $V\gamma\gamma$ production.
Both the shape and normalisation of the total non-resonant background are obtained directly from data,
as described in Section~\ref{sec:param}.
Simulation events for the total non-resonant background are used in Figure~\ref{fig:controls} and for the choice of background analytic parametrisation as described in Section~\ref{sec:param}.
For the production of the resonant SM Higgs boson, events from the $Wh$ and $Zh$ processes
were generated with \pythiaeight with the A14 tune and the NNPDF2.3LO PDF set.
The gluon--gluon fusion (ggF) and vector-boson fusion (VBF) samples were generated with \powheg v2~\cite{Nason:2004rx,Alioli:2008tz,Frixione:2007vw,Nason:2009ai,Alioli:2010xd}
interfaced to \pythiaeight with
the AZNLO~\cite{STDM-2012-23} tune and the CT10 PDF
set~\cite{Lai:2010vv}. Samples of $t\overline{t}h$ events were generated with
\amcnlo\_aMC@NLO 2.2.3 interfaced to \pythiaeight with the NNPDF3.0LO PDF set. Samples of $b\overline{b}h$ events were
generated with \amcnlo\_aMC@NLO 2.2.3 interfaced to \pythiaeight with the A14 tune and the NNPDF2.3LO
PDF set. The non-resonant diphoton processes with associated jets were generated using
\sherpa~\cite{Gleisberg:2008ta}. Matrix elements (ME) were
calculated with up to three partons at LO and merged with the \sherpa parton
shower (PS)~\cite{Schumann:2007mg} using the ME+PS@LO prescription~\cite{Hoeche:2009rj}. The CT10 PDF set was used in
conjunction with a dedicated parton-shower tuning developed by the authors of \sherpa. The $V\gamma$ and $V\gamma\gamma$ samples were
generated using \sherpa with the CT10 PDF set.
 
The cross-sections for the SM Higgs boson processes were
calculated at next-to-leading order (NLO) in electroweak theory and
next-to-next-to-leading order (NNLO) in QCD for the VBF,
$Zh$ and $Wh$ samples~\cite{deFlorian:2016spz,Ciccolini:2007jr,Ciccolini:2007ec,Bolzoni:2010xr,Mimasu:2015nqa,Brein:2003wg,Altenkamp:2012sx,Denner:2011id} and next-to-next-to-next-to-leading order
plus next-to-next-to-leading logarithm (N$^{3}$LO+NNLL) in QCD for the ggF
sample~\cite{deFlorian:2016spz,Anastasiou:2015ema,Anastasiou:2016cez,Actis:2008ug,Anastasiou:2008tj}.
The $t\overline{t}h$ cross-section was calculated with NLO accuracy in QCD with NLO electroweak corrections~\cite{Beenakker:2002nc,Dawson:2003zu,Yu:2014cka,Frixione:2015zaa}.
The $b\overline{b}h$ cross-section was obtained by matching the five-flavor scheme cross section accurate to NNLO in QCD with the four-flavor scheme cross section accurate to NLO in QCD~\cite{Dawson:2003kb,Dittmaier:2003ej,Harlander:2011aa}.
The SM Higgs boson mass was set to 125.09~\GeV~\cite{HIGG-2014-14} and its branching fraction
to decay into two photons was 0.227\%~\cite{deFlorian:2016spz}.
 
Different pile-up conditions from same and neighbouring bunch crossings as a function of the instantaneous luminosity were
simulated by overlaying minimum-bias events, generated with \pythiaeight with the MSTW2008LO PDF set~\cite{Martin:2009iq} and
the A3~\cite{ATL-PHYS-PUB-2016-017} tune, onto all hard-process events.
Differences between the simulated and observed distributions of the number of
interactions per bunch crossing were corrected for by applying weights to simulated events.
Detector effects were simulated using a full simulation~\cite{Aad:2010ah}
performed using \GEANT4~\cite{Agostinelli:2002hh} for the signals, SM Higgs boson
processes, $V\gamma$ and $V\gamma\gamma$ backgrounds. The diphoton continuum background and some of
the signal samples were simulated using a fast simulation of the calorimeter based on \atlfasttwo~\cite{Aad:2010ah}.
 
\section{Event reconstruction}
\label{sec:objsel}

Photons are reconstructed in the region $|\eta| < 2.37$, excluding the EM calorimeter transition region $1.37<|\eta|<1.52$, from clusters of energy deposits in the EM calorimeters.
Clusters without a matching track or reconstructed conversion vertex in the ID are classified as unconverted photons. Those with a matching reconstructed conversion vertex or with a matching track, consistent with originating from a photon conversion, are classified as converted photons. The reconstruction efficiency is 99\% for photons and the conversion reconstruction efficiency is 70\%~\cite{EGAM-2018-01}.
The photon energy is calibrated using a multivariate regression algorithm
trained with fully reconstructed MC samples and then corrected using data-driven techniques~\cite{EGAM-2018-01}.
The overall energy scale in data and the difference in the constant term on the
resolution between data and simulation are estimated from using a sample with $Z$ boson decays into electrons~\cite{EGAM-2018-01}.
The photon direction is estimated using either EM calorimeter longitudinal
segmentation (if unconverted) or the conversion vertex position (if converted), together with constraints from
the $pp$ collision point.
 
To reduce the misidentification of hadronic jets containing a high-$\pt$ neutral hadron (e.g. $\pi^0$) decaying into two photons,
`Tight' identification criteria~\cite{EGAM-2018-01} are applied.
The photon identification is based on the lateral profile of the energy
deposits in the first and second layers of the EM calorimeter, and
on the shower leakage fraction in the hadronic calorimeter.
The selection requirements are tuned for converted and unconverted photon candidates, separately.
The identification efficiency for unconverted and converted photons ranges from 85\% to 99\%
between 30\,\GeV\ and 250\,\GeV~\cite{EGAM-2018-01}. Corrections are
applied to the EM shower-shape variables for simulated photons, to account for small differences between data and
simulation.
 
To further suppress hadronic backgrounds, requirements on two photon isolation variables are applied.
The first variable, $\et^\text{iso}$, calculates the sum of the transverse energies
deposited in topological clusters~\cite{Aad:2016upy} in the calorimeter within a cone of size ${\Delta R = 0.2}$ around each photon.
The photon cluster energy and an estimate of the energy deposited by the photon
outside its associated cluster are also subtracted from this sum.
To reduce underlying-event and pile-up effects, $\et^\text{iso}$ is
further corrected using the method described in Refs.~\cite{STDM-2010-08,Cacciari:2008gn,Cacciari:2009dp}.
The second variable expresses track-based isolation, defined as the scalar sum of the transverse momenta
of all tracks with $\pT>1$~\GeV\ and consistent with originating from the
primary vertex (PV) within a cone of size ${\Delta R = 0.2}$ around each photon. The isolation efficiency for photons, which is mostly independent of their kinematic variables, is about
90\%.

Events are required to have at least one PV, defined as a vertex associated
with at least two tracks with $\pt > 0.5~\GeV$. In each event, the PV most likely to be the origin of the diphoton,
selected from the PV candidates using a neural network~\cite{HIGG-2016-21}, is required to be consistent with the PV with
the highest sum of squared transverse momenta of associated tracks.
The neural network algorithm selects a diphoton vertex within 0.3 mm of the true $h \rightarrow \gamma\gamma$ production vertex in 79\% of simulated gluon–gluon fusion events.
For the other Higgs production modes this fraction ranges from 84\% to 97\%, increasing with jet activity or the presence of charged leptons~\cite{HIGG-2016-21}.

Electrons are reconstructed from energy deposits measured in the EM calorimeter
that are matched to tracks from ID~\cite{EGAM-2018-01}. They are required to satisfy $|\eta| < 2.47$, excluding the EM calorimeter
transition region $1.37 < |\eta| < 1.52$, and to have $\pT > 10$ \GeV. The electrons are identified using
a likelihood-based algorithm that uses track and shower-shape variables. The `MediumLLH'
criteria are applied, providing an identification efficiency varying from 85\% to 95\% as a function of \ET~\cite{EGAM-2018-01}.
Loose calorimeter and track isolation requirements are applied to electrons. The efficiency of the isolation
requirements is 98\%~\cite{Aad:2014lwa}.

Muons are reconstructed from high-quality track segments in the muon spectrometer. In the region $|\eta| < 2.5$, they must be matched to ID tracks. They are required to have $\pT > 10$~\GeV\ and $|\eta| < 2.7$.
The muon `medium' criteria are applied with a 96\%~\cite{Aad:2016jkr} identification efficiency. The muon
candidates must also satisfy loose calorimeter and track isolation criteria. The
combined isolation efficiency varies from 95\% to 99\% as a
function of \pT from 25~\GeV\ to 60~\GeV~\cite{Aad:2016jkr}.
 
The significance of the track's transverse impact parameter relative to the PV is required to be $|d_0|/\sigma_{d_0} <5\ (3)$ for electrons (muons). The longitudinal
impact parameter $z_0$ must satisfy $|z_0|\sin\theta <$ 0.5 mm for electrons and muons.
 
Jets are reconstructed from three-dimensional topological clusters using the anti-\kT algorithm~\cite{Cacciari:2008gp,Fastjet}
with a radius parameter of $R = 0.4$. The jets are required to have $\pT > 20$~\GeV\ and $\abseta < 4.5$ for
the \etmiss calculation and $\pT > 25$~\GeV\ and $\abseta < 4.4 $ for the event selection. Jets with $\abseta<2.4$
and $\pT < 60$~\GeV\ must satisfy the jet vertex tagger (JVT)
selection~\cite{ATL-PHYS-PUB-2014-001}, in which a jet is identified as originating from the PV
depending on a likelihood value calculated from the track information. In addition, quality criteria
are applied to the jets, and events with jets consistent with noise in the calorimeter or non-collision backgrounds are
rejected~\cite{ATLAS-CONF-2015-029}.
 
Reconstruction ambiguities between photons, electrons, muons, and jets are resolved using an `overlap removal' procedure among all the objects in the following order.
First, electrons, muons, and jets found within $\Delta R = 0.4$ of a photon are removed. Next, jets found within $\Delta R = 0.2$ of an electron are removed. Lastly, electrons and muons within $\Delta R = 0.4$ of the remaining jets are removed.
A different overlap removal strategy was used in the previous study~\cite{SUSY-2017-01} and the selection is discussed in Section~\ref{sec:excessevt}.
It was motivated by the prioritisation of electrons, as opposed to photons.
The results show no significant difference in sensitivity between these two strategies.
 
Jets containing a $b$-hadron are identified using the MV2c10~\cite{FTAG-2018-01,ATL-PHYS-PUB-2017-013}
multivariate discriminant built with information from track impact parameters and the presence of reconstructed secondary vertices,
which applies a multi-vertex fitter to reconstruct the hadron decay chain $b \to c$.
A value of the discriminating variable is chosen such that it provides a $b$-tagging efficiency of 70\% in simulated $\ttbar$ events.
The rejection for $c$-jets and jets originating from gluons or light ($u$, $d$, $s$) quarks are 8.9 and 300~\cite{FTAG-2018-01}, respectively.
An additional energy correction is applied to $b$-jets to account for the presence of muons in the jet~\cite{FTAG-2018-01}.

The $\met$ is calculated as the magnitude of the negative vectorial sum of the transverse momenta
of calibrated photons, electrons, muons and jets associated with the PV\@. The transverse
momenta of all remaining tracks that originate from the PV
but are not already used in the \etmiss calculation are summed and taken into account in the \etmiss calculation.
This term is defined as
the track-based soft term~\cite{PERF-2016-07}.
In this way, the \etmiss is adjusted for the best calibration of the jets and the other identified physics objects above, while maintaining pileup independence in the soft term.
 
\section{Event selection}
\label{sec:evtsel}
\subsection{Baseline selection}
 
Each event is first required to contain at least two photons with $\pT > 22$~\GeV.
The photons are ordered by their $\pT$. The leading and subleading photons are then required to have $\pT^{\gamma}/m_{\gamma\gamma} > 0.35$ and $0.25$, respectively, where $m_{\gamma\gamma}$ is the invariant mass of the leading and subleading photon pair.
The signal region is defined as $105 < m_{\gamma\gamma} < 160$~\GeV, where $m_{\gamma\gamma}$ is calculated using the photon momentum vectors recomputed relative to the PV.
The selected events are divided into 12 categories based on the number of leptons ($N_{\ell}$), number of jets ($N_{j}$), the invariant mass of the two highest-$\pt$ jets ($m_{jj}$), and the \etmiss significance $\setmiss = \etmiss / \sqrt{\sum E_\mathrm{T}}$. The total transverse energy $\sum E_\mathrm{T}$ is calculated from the scalar sum of the transverse momenta of the calibrated photons, electrons, muons and jets used in the \etmiss calculation described in Section~\ref{sec:objsel}, as well as the tracks not associated with these but consistent with originating from the PV\@.
Because both the $\etmiss$ and $\sum E_\mathrm{T}$ resolutions increase linearly with the number of pileup events, $\setmiss$ is more resilient to pileup than $\etmiss$.
No $b$-jet veto is applied in the baseline selection.
The 12 categories are defined in Table~\ref{tab:cat_final}.
The \cc\nnn signal sample with $m(\cc/\nnn)=150~\GeV$ and $m(\nn)=0.5~\GeV$ is used to optimise the boundary of each category to maximise the significance when combining all 12 categories.
This signal point has low $\met$, where the diphoton channel is expected to have a better sensitivity than the channel with the SM Higgs boson decaying into a pair of $b$-quarks~\cite{Aad:2019vvf,Sirunyan:2018ubx}.
The `Leptonic' and `Hadronic' categories are used to accommodate the most clearly identifiable leptonic and hadronic decays of the $W$ boson, while the `Rest' category retains all additional signal topologies.
The signal $\cc\nnn \to W^{\pm}\nn h \nn$ has the highest expected significance in the Leptonic categories, and the $h\tilde{G} h\tilde{G}$ signals have the highest expected significance in the Rest categories.
Because the different signal models and mass points have different \pt distributions as shown in Figure~\ref{fig:genplot}, and since \pt and \setmiss distributions are highly correlated,
each region is divided into \setmiss bins to improve the sensitivity.
The regions do not change significantly if a different mass point is used for optimisation.

\begin{table*}[!htp]
\caption{Criteria used in the categorisation.}
\begin{center}
\begin{tabular}{ccc}
\toprule\hline
Channels  &           Names  & Selection \\
\hline
& Category 1 & $0<S_{\etmiss} \leq 2$, $N_{\ell} \geq 1 $ \\
& Category 2 & $2<S_{\etmiss} \leq 4$, $N_{\ell} \geq 1 $ \\
Leptonic           & Category 3 & $4<S_{\etmiss} \leq 6$, $N_{\ell} \geq 1 $ \\
& Category 4 & $S_{\etmiss} > 6$, $N_{\ell} \geq 1 $ \\
\hline
&  Category 5 & $5<S_{\etmiss} \leq 6$, $N_{\ell} = 0 $,  $N_{j} \geq 2 $, $m_{jj} \in [40,120]~\GeV$\\
&  Category 6 & $6<S_{\etmiss} \leq 7$, $N_{\ell} = 0 $,  $N_{j} \geq 2 $, $m_{jj} \in [40,120]~\GeV$\\
Hadronic          &  Category 7 & $7<S_{\etmiss} \leq 8$, $N_{\ell} = 0 $,  $N_{j} \geq 2 $, $m_{jj} \in [40,120]~\GeV$\\
&  Category 8 & $S_{\etmiss} > 8$, $N_{\ell} = 0 $,  $N_{j} \geq 2 $, $m_{jj} \in [40,120]~\GeV$\\
\hline
&  Category 9 & $6<S_{\etmiss} \leq 7$, $N_{\ell} = 0 $,  $N_{j} < 2 $ or ( $N_{j} \geq 2$, $m_{jj} \notin [40,120]~\GeV$)\\
&  Category 10 & $7<S_{\etmiss} \leq 8$, $N_{\ell} = 0 $, $N_{j} < 2 $ or ( $N_{j} \geq 2$, $m_{jj} \notin [40,120]~\GeV$)\\
Rest		    &  Category 11 & $8<S_{\etmiss} \leq 9$, $N_{\ell} = 0 $, $N_{j} < 2 $ or ( $N_{j} \geq 2$, $m_{jj} \notin [40,120]~\GeV$)\\
&  Category 12 & $S_{\etmiss} > 9$, $N_{\ell} = 0 $, $N_{j} < 2 $ or ( $N_{j} \geq 2$, $m_{jj} \notin [40,120]~\GeV$)\\
\hline
\bottomrule
\end{tabular}
\label{tab:cat_final}
\end{center}
\end{table*}
 
Figure~\ref{fig:controls} shows the distribution of \setmiss after the selection of diphoton candidates
with $120 < m_{\gamma\gamma} < 130$~\GeV, where signal dominates.
The shapes and normalisations of the $V\gamma$ and $V\gamma\gamma$ contributions are obtained from the MC simulation. The shape of the $\gamma\gamma$ contribution is obtained from the MC simulation while the normalisation is fixed to the yields in the sidebands ($105 < m_{\gamma\gamma} \leq 120$~\GeV, $130 \leq m_{\gamma\gamma} < 160$~\GeV) of the data multiplied by the diphoton purity among all the backgrounds. The diphoton purity is measured in the data, using a two-dimensional sideband technique by counting the number of events in which one or both photons satisfy or fail to satisfy  the identification or isolation requirements~\cite{Aad:2012tba}. The diphoton purity varies from 65\% to 93\% for different categories. The shape of the $\gamma$+jets contribution is obtained using the data distribution in a control region where the event selection is the same as for the signal region but one of the photons fails to satisfy the identification criteria, after subtracting the contamination from $\gamma\gamma$, $V\gamma$ and $V\gamma\gamma$ using MC simulation. Its normalisation is fixed to the $\gamma$+jets purity and varies from 34\% to 7\% of the total yield in different categories.
 
\begin{figure}[!htpb]
\centering
\begin{tabular}{cccc}
\includegraphics[width=0.49\textwidth]{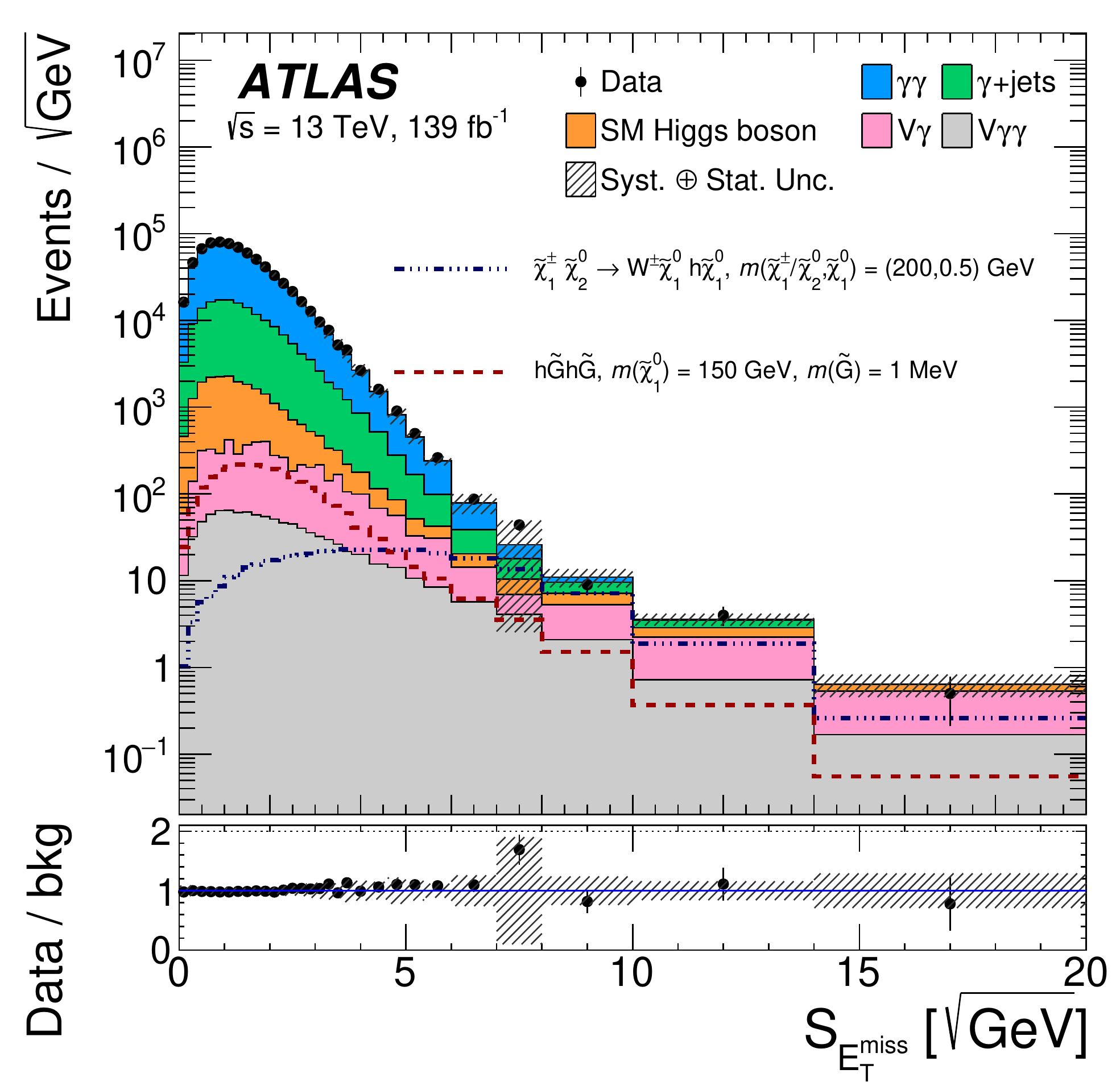}
\end{tabular}
\caption{
The distribution of \setmiss after the selection of diphoton candidates
with $120 < m_{\gamma\gamma} < 130$~\GeV.
Expected distributions are shown for the $\cc\nnn \to W^{\pm} \nn h \nn $ signal with $m(\cc/\nnn)=200~\GeV$ and $m(\nn)=0.5~\GeV$,
and the $h\tilde{G} h\tilde{G}$ signal with $m(\nn)=150~\GeV$ and $m(\tilde{G})=1~\MeV$.
These overlaid signal points are representative of the model kinematics.
The sum in quadrature of the MC statistical and experimental systematic uncertainties in the total background is shown as the hatched bands,
while the theoretical uncertainties in the background normalisation are not included. The $tt\gamma$ and $tt\gamma\gamma$ processes have a negligible contribution and are not represented.
Overflow events are included in the rightmost bin. The lower panel shows the ratio of the data to the background, called ``bkg''.
}
\label{fig:controls}
\end{figure}
 
\subsection{Follow-up selection}
\label{sec:excessevt}
To check the small excess of events observed in the previous search from ATLAS
using 36.1 \ifb\ of \(pp\) collision data~\cite{SUSY-2017-01}, two signal regions (`SR1L$\gamma\gamma$-a' and `SR1L$\gamma\gamma$-b') defined in the previous search are reused in this analysis.
Events are required to have exactly one lepton with $\pt > 25$~\GeV\ and exactly two photons with $\pt >$ 40 (30)~\GeV\ for the leading (subleading) photon.
The invariant mass of the two photons is required to be $105 < \mgg < 160$~\GeV,
with $\met>40$~\GeV.
The difference in azimuthal angle in the transverse plane between the diphoton system and the lepton plus $\met$ vector is required to be greater than 2.25 radians.
To reduce contributions from $t\bar{t}h$, a $b$-jet veto is used in both the signal regions.
 
To further reduce  contributions from SM backgrounds, the transverse mass $m^{W}_{\mathrm{T}}$~\cite{SUSY-2017-01} of the lepton and $\met$,
and the three-body transverse mass $m^{W \gamma i}_{\mathrm{T}}$~\cite{SUSY-2017-01} of the lepton, $\met$ and the $i^{th}$ photon ordered by $\pt$
are used to define the two orthogonal signal regions.
For both signal regions, events are required to have $m^{W \gamma 1}_{\mathrm{T}} > 150$~\GeV\ and $m^{W \gamma 2}_{\mathrm{T}} > 80$~\GeV.
The first signal region, `SR1L$\gamma\gamma$-a', selects events with $m^{W}_{\mathrm{T}}>110$~\GeV\ and $m^{W \gamma 2}_{\mathrm{T}}>140$~\GeV\ while the
events that fail to satisfy these requirements define the second signal region (`SR1L$\gamma\gamma$-b').
 
\section{Signal and background parameterisation}
\label{sec:param}
The signals and the SM Higgs boson background mass distributions are described independently using double-sided
Crystal Ball functions (as defined in Ref.~\cite{HIGG-2014-04}). The parameter values for the functions are extracted by fitting the
diphoton invariant mass distributions of the MC simulation for each category.
The expected normalisations are calculated from the theoretical cross-sections multiplied by the acceptance and efficiency from the MC simulation.
 
The normalisation and shape of the non-resonant background are extracted by fitting the diphoton invariant mass distribution in data for each category.
Following the method used in the measurement of the SM Higgs boson decaying into two photons~\cite{Aad:2014eha},
several candidate analytic functions are chosen for the non-resonant background parameterisation:
the exponential functions of different-order polynomials, Bernstein polynomials of
different order, and an adapted dijet function~\cite{Aaltonen:2008dn}. The potential bias, denoted by $\Delta N^\textrm{non-res}_\mathrm{bkg}$,
from the functional form modelling the continuum background in each category is estimated.
It is defined as the maximal signal yield extracted from the fit to a continuum-background-only diphoton invariant mass distribution.
This distribution is taken from MC simulations and is normalised to the integrated luminosity of 139~\ifb,
with small statistical uncertainty, using a signal-plus-background model. The Higgs boson mass varies from 115~\GeV\ to 135~\GeV~\cite{Aad:2014eha}.
This is to ensure the bias from choosing different background models is conservatively estimated.
For categories 2 to 12, the functional form with $\Delta N^\textrm{non-res}_\mathrm{bkg}$ less than $20\%$ of the statistical uncertainty in data and with the fewest free parameters is chosen as the nominal background function. In the case of Category 1, with large MC statistical uncertainty, none of the functional forms satisfies the criterion on the fraction of the statistical uncertainty in data, thus the functional form with the smallest $\Delta N^\textrm{non-res}_\mathrm{bkg}$ is chosen.
The $\Delta N^\textrm{non-res}_\mathrm{bkg}$ value of the chosen functional form is taken as the non-resonant background modelling
uncertainty in each category and is shown in Table~\ref{tab:decision}.

\begin{table*}[!htbp]
\centering
\caption{The analytic functions used to model the non-resonant background, the extracted signals from the background-only fits ($\Delta N^\textrm{non-res}_\mathrm{bkg}$) to the MC and the relative uncertainty in the non-resonant background within $120 < m_{\gamma\gamma} < 130$~\GeV\ ($\Delta N^\textrm{non-res}_\mathrm{bkg}$/$N^{\textrm{non-res.}}_{\mathrm{bkg}}$) for each category. The variable $x$ is defined as $m_{\gamma\gamma}/\sqrt{s}$ while $a$ and $b$ are parameters of the background functions. The $C_3^j$ are binomial coefficients and the $b_{j,3}$ are the fitted parameters for the third order Bernstein polynomial parameterization. }
\begin{tabular}{llcc}
\toprule
\toprule
\multicolumn{1}{l}{ Category } &  \multicolumn{1}{l}{ Function}  &  \multicolumn{1}{c}{ $\Delta N^\textrm{non-res}_\mathrm{bkg}$ } &
\multicolumn{1}{c}{ $\Delta N^\textrm{non-res}_\mathrm{bkg}$/$N^{\textrm{non-res.}}_{\mathrm{bkg}}$ [\%]} \\
\midrule
1   &   $(1-x^{1/3})^b \cdot x^a$                   & 5.5 & 2.4 \\
2   &   $\sum_{j=0}^3 C_3^j x^j(1-x)^{3-j} b_{j,3}$ & 1.8 & 2.4 \\
3   &   $\exp(a\cdot x)$                            & 0.6 & 3.6 \\
4   &   $\exp(a\cdot x)$                            & 0.3 & 3.7 \\
5   &   $\exp(a\cdot x)$                            & 1.6 & 2.8 \\
6   &   $\exp(a\cdot x)$                            & 0.5 & 3.3 \\
7   &   $\exp(a\cdot x)$                            & 0.3 & 5.1 \\
8   &   $\exp(a\cdot x)$                            & 0.2 & 4.6 \\
9   &   $\exp(a\cdot x)$                            & 1.5 & 2.3 \\
10  &   $\exp(a\cdot x)$                            & 0.6 & 2.5 \\
11  &   $\exp(a\cdot x)$                            & 0.4 & 5.6 \\
12  &   $\exp(a\cdot x)$                            & 0.4 & 3.0 \\
\bottomrule
\bottomrule
\end{tabular}
\label{tab:decision}
\end{table*}
 
 
\section{Systematic uncertainties}
\label{sec:sys}
Uncertainties from experimental and theoretical sources that affect the signal
efficiency and the SM Higgs boson background yield are estimated from the MC simultation.
The non-resonant background is obtained directly from the fit to the data.
The only systematic uncertainty in the non-resonant background is the potential
bias in $\Delta N^\textrm{non-res}_\mathrm{bkg}$ from the choice of background modelling.
A summary of the experimental and theoretical uncertainties in the yield from the SM Higgs boson background processes, non-resonant background, and
signal production is shown in Table~\ref{tab:unc}.
 
\begin{table*}[!htp]
\caption{Breakdown of the dominant systematic uncertainties.
The uncertainties (in \%) in the yield of signals, the background
from the SM Higgs boson processes and non-resonant background are shown.
All production modes of the SM Higgs boson are considered together. 
A ``--'' indicates that the systematic uncertainty is not applicable to the corresponding sample.
If a given source has a different
impact on the various categories, the given
range corresponds to the smallest and largest impacts
among categories or among the different signal models used
in the analysis. In addition, the potential bias coming from non-resonant
background modelling is shown relative to the background in the signal region $120 < m_{\gamma\gamma} < 130$~\GeV.
}
\begin{center}
\begin{tabular}{lccc}
\toprule
\toprule
\multirow{3}{*}{Source}  & \multirow{3}{*}{Signals [\%]}  & \multicolumn{2}{c}{Backgrounds [\%]} \\
\cline{3-4}
&   & SM Higgs boson & \begin{tabular}{c} Non-resonant \\  background \end{tabular} \\
\midrule
\multicolumn{3}{l}{Experimental} \\
\midrule
Luminosity     & \multicolumn{2}{c}{1.7} & --   \\  
Jets (scale, resolution, JVT)        & 0.2--3.3 & ~~0.9--31 &   --    \\
Electron/Photon (scale, resolution) & 0.3--1.5 & 0.6--2.7 &  --   \\
Photon (identification, isolation, trigger)   & 2.2--2.6 & 2.8--4.3  &  --   \\
Electron (identification isolation) & 0.0--0.5 & 0.0--0.6  &   --   \\
Muon (identification, isolation, scale, resolution) & $<0.6$ & $<0.3$  &   -- \\
\etmiss\ reconstruction (jets, soft term) & $<0.7$ & ~~0.4--14 &  --   \\
Pile-up reweighting & 0.3--1.8 & 1.3--1.5 &  -- \\
Non-resonant background modelling &  \multicolumn{2}{c}{--} & 2--6  \\
\midrule
\multicolumn{3}{l}{Theoretical} \\
\midrule
Factorisation and renormalisation scale & $<1$ & 4.1--6.5 &  --  \\
PDF+$\alpha_\mathrm{S}$ & $<6.6$  & 3.3--6.4 &  --  \\
Multiple parton--parton interactions & \multicolumn{2}{c}{$< 1$}  &  --   \\
$\cal{B}$($H\rightarrow \gamma\gamma$) & \multicolumn{2}{c}{1.73} &  --  \\
\bottomrule
\bottomrule
\end{tabular}
 
\label{tab:unc}
\end{center}
\end{table*}
 
The uncertainty in the combined 2015--2018 integrated luminosity is 1.7\%~\cite{ATLAS-CONF-2019-021}, obtained using the LUCID-2 detector~\cite{LUCID2} for the primary luminosity measurements.
 
The efficiency of the diphoton trigger used to select events is evaluated in MC simulation using
a trigger matching technique and in data using a bootstrap method~\cite{Aad:2019wsl}.
The uncertainty in the trigger efficiency for events with $105 < m_{\gamma\gamma} < 160$~\GeV\ is found to be $0.4\%$.
 
The uncertainty in the vertex selection efficiency is assessed by comparing the efficiency of finding
photon-pointing vertices in $Z \rightarrow e^+e^-$ events in data with that in MC simulation~\cite{Aaboud:2016tru}. The resulting uncertainty is found
to be negligible in the inclusive photon selection.

The systematic uncertainties due to the photon energy scale and
resolution are obtained from Ref.~\cite{EGAM-2018-01}.
The uncertainty in the energy scale has an effect below $1\%$ on the normalisation of the signals and the SM Higgs boson background in the \pT range of the photons used in the analysis.
The uncertainty in the energy resolution has an effect below $2\%$ on the normalisation of the signals and the SM Higgs boson background. The uncertainties affecting the signal and the SM Higgs boson background mass distributions due to the photon energy scale and resolution are also evaluated.
The uncertainties vary from below 1\% to 20\% for different categories and for different SM Higgs boson production processes. Overall, they amount to less than 3\% of the total SM Higgs boson background.
 
Uncertainties in photon identification and isolation efficiencies are estimated~\cite{EGAM-2018-01}, and their impact on the number of events in each category is quantified.
The photon identification uncertainty varies in the range $1\%$--$3\%$ for the SM Higgs boson background and $1\%$--$2\%$ for the signals in all categories.
The uncertainty in the photon calorimeter isolation efficiency is calculated from efficiency differences
between applying and not applying corrections derived from inclusive photon events to the isolation variables in simulation.
The measurements of the efficiency correction factors using inclusive photon
events are used to derive the uncertainty in the photon track isolation efficiency.
The photon isolation efficiency uncertainty is found to be in the range $1\%$--$3\%$ for the SM Higgs boson background and $1\%$--$2\%$ for the signals.
 
Migration of events among categories occurs if the energies of identified particles, jets and the \etmiss, are varied within their uncertainties.
The uncertainties in the jet energy scale, resolution~\cite{PERF-2016-04} and jet vertex
tagger are propagated to the \etmiss calculation. In addition, the
uncertainties in the scale and resolution of the \etmiss soft term are estimated by using the method
described in Ref.~\cite{PERF-2016-07}.
The overall jet and \etmiss uncertainties in the SM Higgs boson processes vary from 1.0\% to 34\% for each category
and for different SM Higgs boson production processes. Overall, they amount to 0.4\%--14\% for the total SM Higgs boson background.
For the signal processes, the overall jet and \etmiss uncertainties range from 0.2\% to 3.3\%.
An uncertainty in the pile-up modelling in MC simulation is accounted for.
This results in an uncertainty of 0.3\%--1.8\% in the signal yield and 1.3\%--1.5\% in the SM Higgs boson yield.
The uncertainties related to the $b$-tagging of jets are typically less than 1.5\% in the SM Higgs boson yield used in the `follow-up' analysis.

The predicted cross-sections of the SM Higgs boson and signal processes are affected by uncertainties due to missing
higher-order terms in perturbative QCD\@. These uncertainties are estimated by varying the factorisation and renormalisation scales up and
down from their nominal values by a factor of two, recalculating the cross-section in each case, and taking the largest deviation from
the nominal cross-section as the uncertainty. The acceptance uncertainty related to the renormalisation and factorisation scales
is less than 1\% for the signal and 3.7\%--5.9\% for the SM Higgs boson processes~\cite{deFlorian:2016spz}.
The normalisation uncertainty of the SM Higgs boson processes is 1.7\% to 2.8\%. For the signal processes, the effect of PDF and $\alpha_\mathrm{S}$
uncertainties in the acceptance times selection efficiency is below 6.6\%.
It is estimated by using the recommendations
of PDF4LHC~\cite{deFlorian:2016spz}. Both the intra-PDF and inter-PDF uncertainties are extracted.
Intra-PDF uncertainties are obtained by varying the parameters of the NNPDF3.0LO PDF set, while inter-PDF uncertainties
are estimated by using alternative PDF sets (CT14~\cite{Dulat:2015mca} at LO and MMHT2014~\cite{Harland-Lang:2014zoa} at LO).
The final inter-PDF uncertainty is the maximum deviation among all the variations from the central
value obtained using the NNPDF3.0LO PDF set.
In the case of the SM Higgs boson processes, the acceptance effect of $\alpha_\mathrm{S}$
and the choice of PDFs ranges from 2.1\% to 2.9\%, and its normalisation effect is 2.5\% to 5.7\%.
The uncertainty in the branching fraction of $h\rightarrow \gamma\gamma$ is 1.73\%~\cite{deFlorian:2016spz}.
The uncertainty in the effect of multiple parton--parton interactions is estimated by switching them on and off in \pythia\
in the production of the ggF SM Higgs boson and signal samples. The resulting uncertainty in the number of events in this sample
conservatively reaches $1\%$ for all the categories.
 
\section{Results}
\label{sec:results}
The results are derived from an unbinned likelihood fit to the \mgg
distributions in the range $105 < \mgg < 160~\GeV$ in each category simultaneously.
The impact of the SM Higgs boson mass uncertainty is negligible.
The signal strength and the background shape parameters are free parameters. The SM Higgs boson yields are taken from the SM predictions as discussed in Section~\ref{sec:datamc}.
The systematic uncertainty in
each nuisance parameter is taken into account by multiplying the likelihood by a Gaussian penalty function centred on the nominal value of this parameter with
a width set to its uncertainty. The nominal value of each SM Higgs boson background nuisance parameter (including its yield) is taken from the simulation normalised to the SM theoretical predictions.

\begin{figure}[!htbp]
\centering
\subfigure[]{ \includegraphics[width=0.45\textwidth]{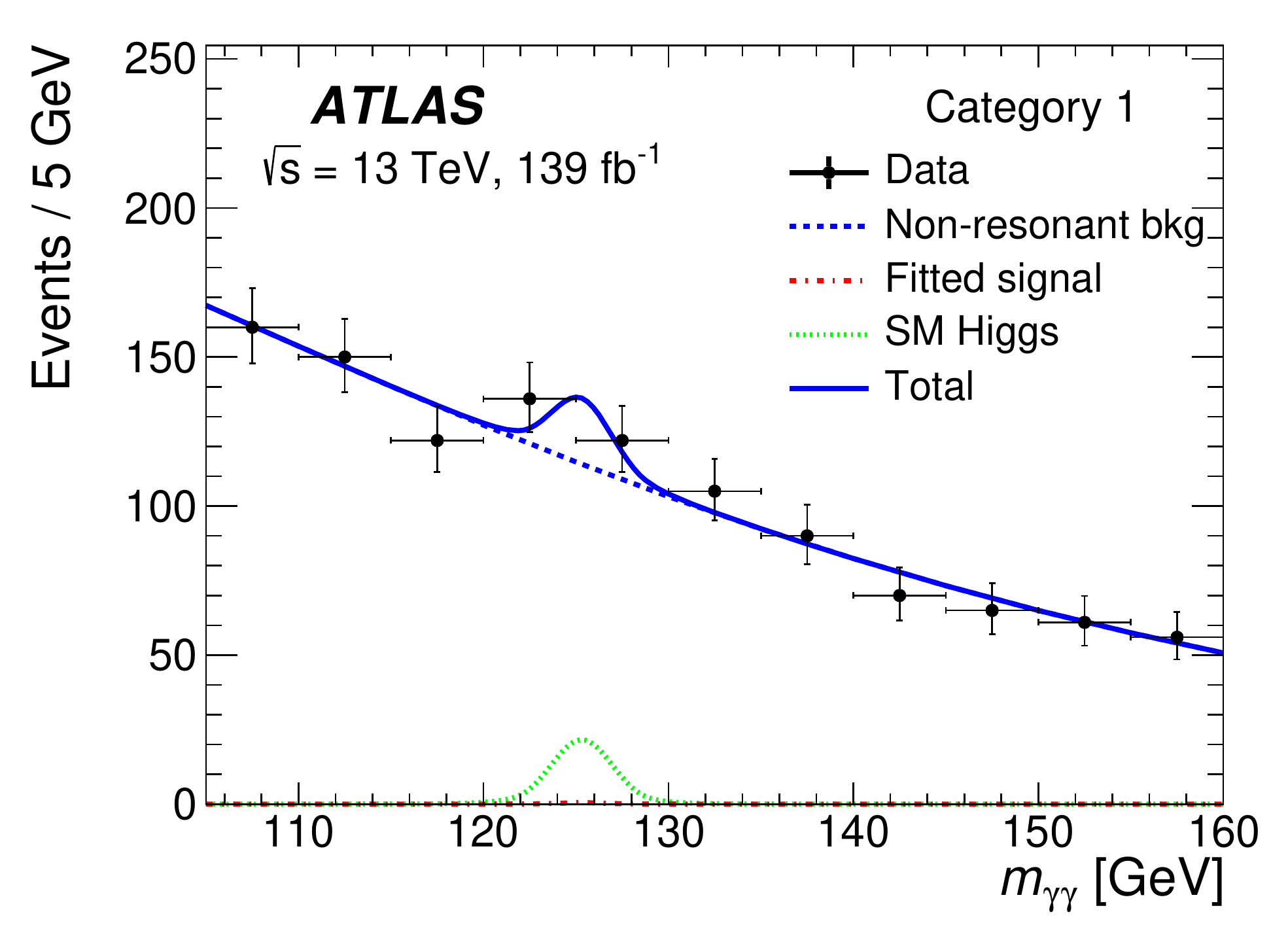}}
\subfigure[]{ \includegraphics[width=0.45\textwidth]{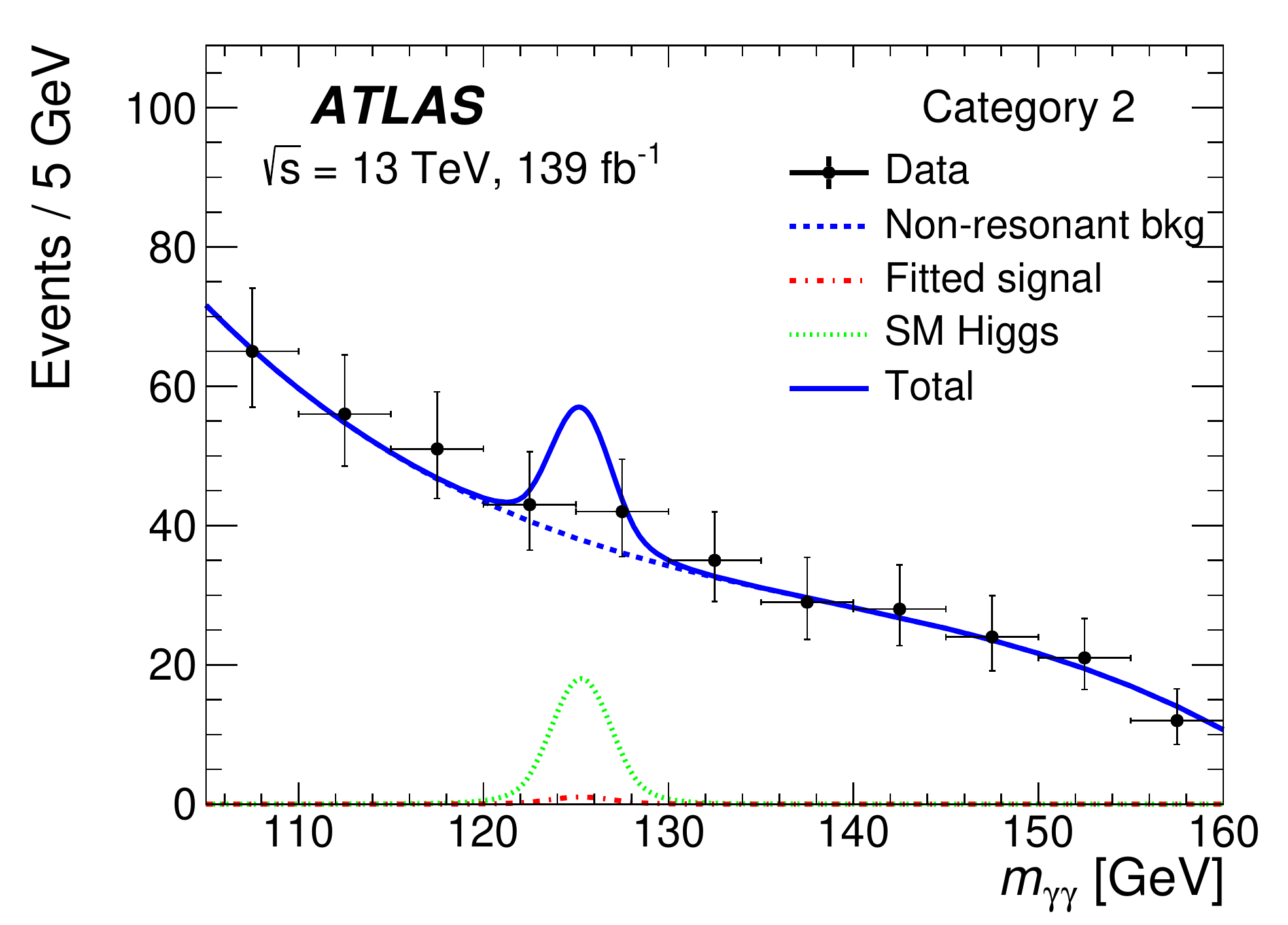}}\\
\subfigure[]{   \includegraphics[width=0.45\textwidth]{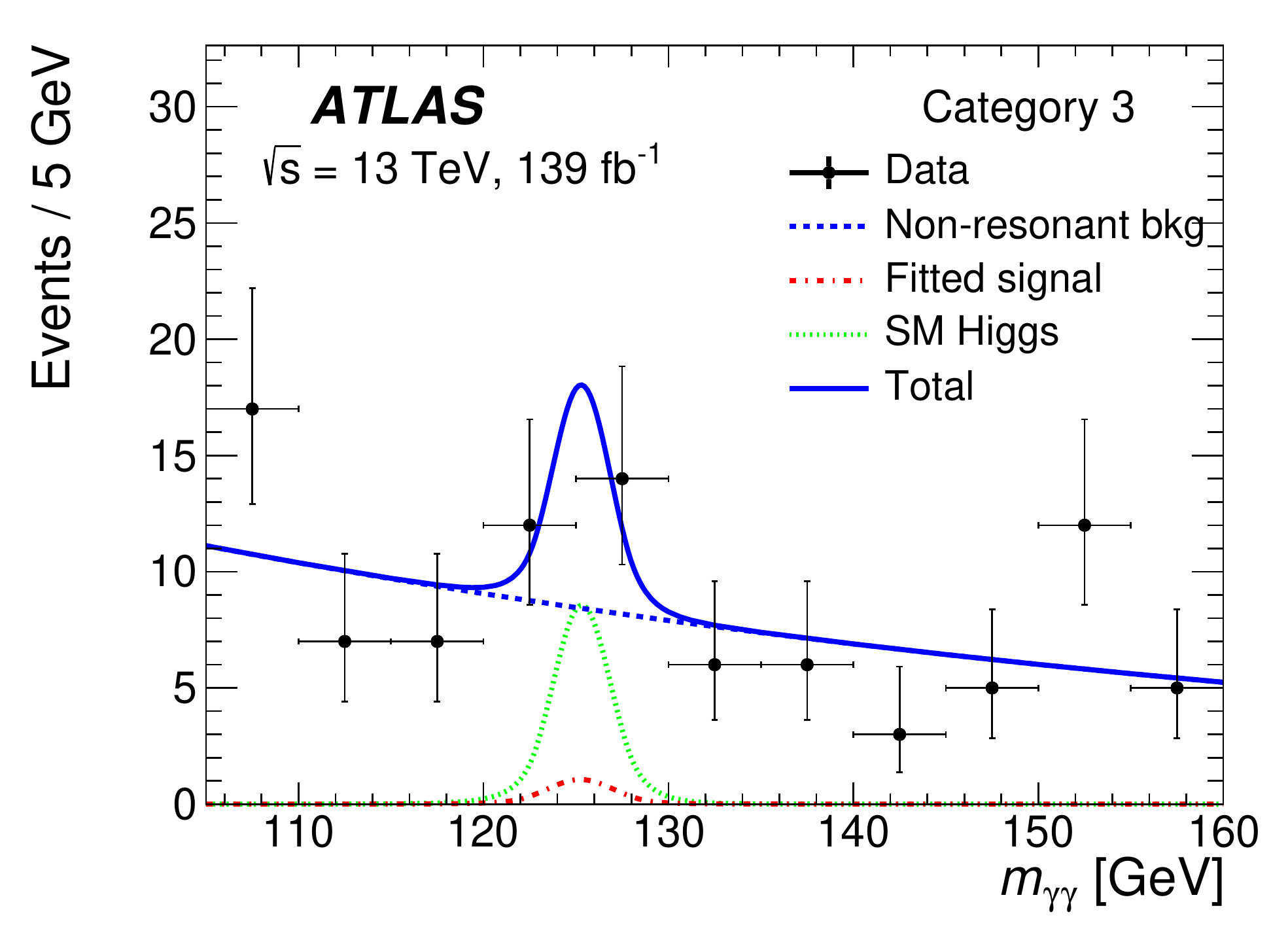}}
\subfigure[]{\includegraphics[width=0.45\textwidth]{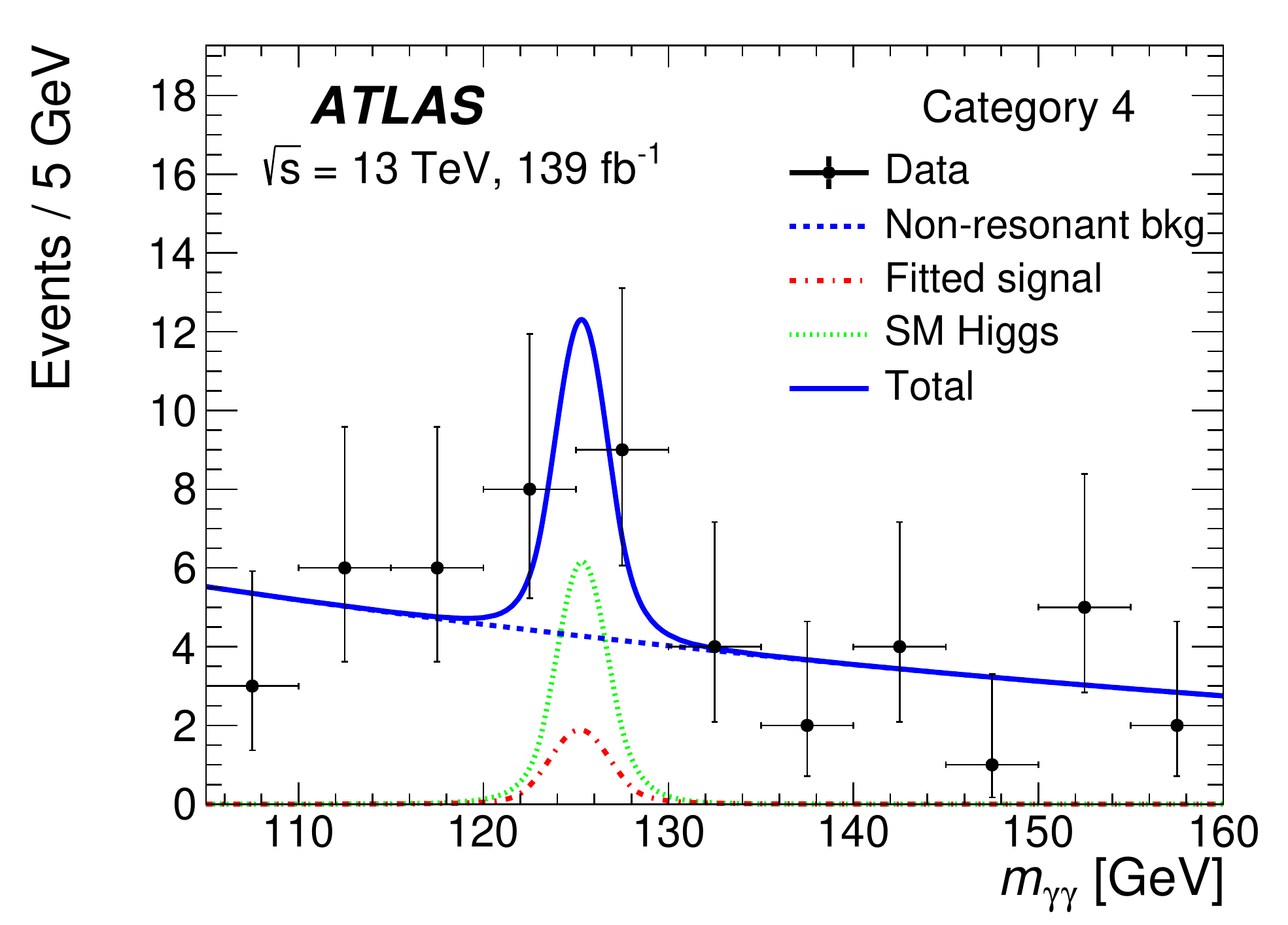}}
\caption{Diphoton invariant mass spectra and the corresponding fitted signal and background in the Leptonic categories (a) 1, (b) 2, (c) 3, and (d) 4.
The signal samples shown correspond to the $\cc\nnn \to W^{\pm}\nn h \nn$ signal with $m(\cc/\nnn)=200~\GeV$ and $m(\nn)=0.5~\GeV$.
The non-resonant background (dashed curve), the SM Higgs boson (dotted curve), and the signal (dash-dotted curve) are obtained
from a simultaneous signal-plus-background fit to the full $m_{\gamma\gamma}$ spectrum for the 12 categories. The total of these contributions is shown by the solid curves.
}
\label{fig:fit_data1}
\end{figure}
 
\begin{figure}[!htbp]
\centering
\subfigure[]{ \includegraphics[width=0.45\textwidth]{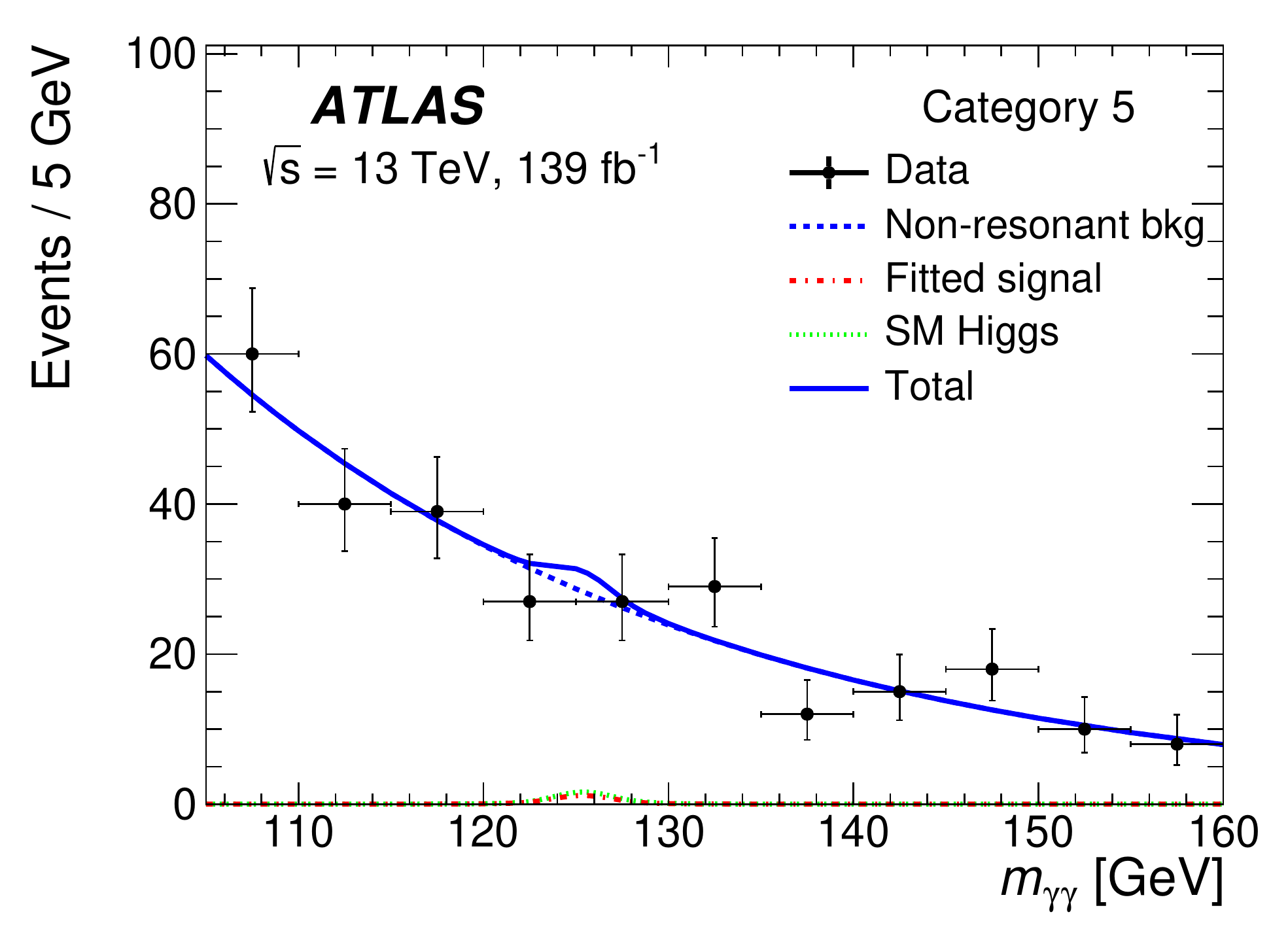}}
\subfigure[]{ \includegraphics[width=0.45\textwidth]{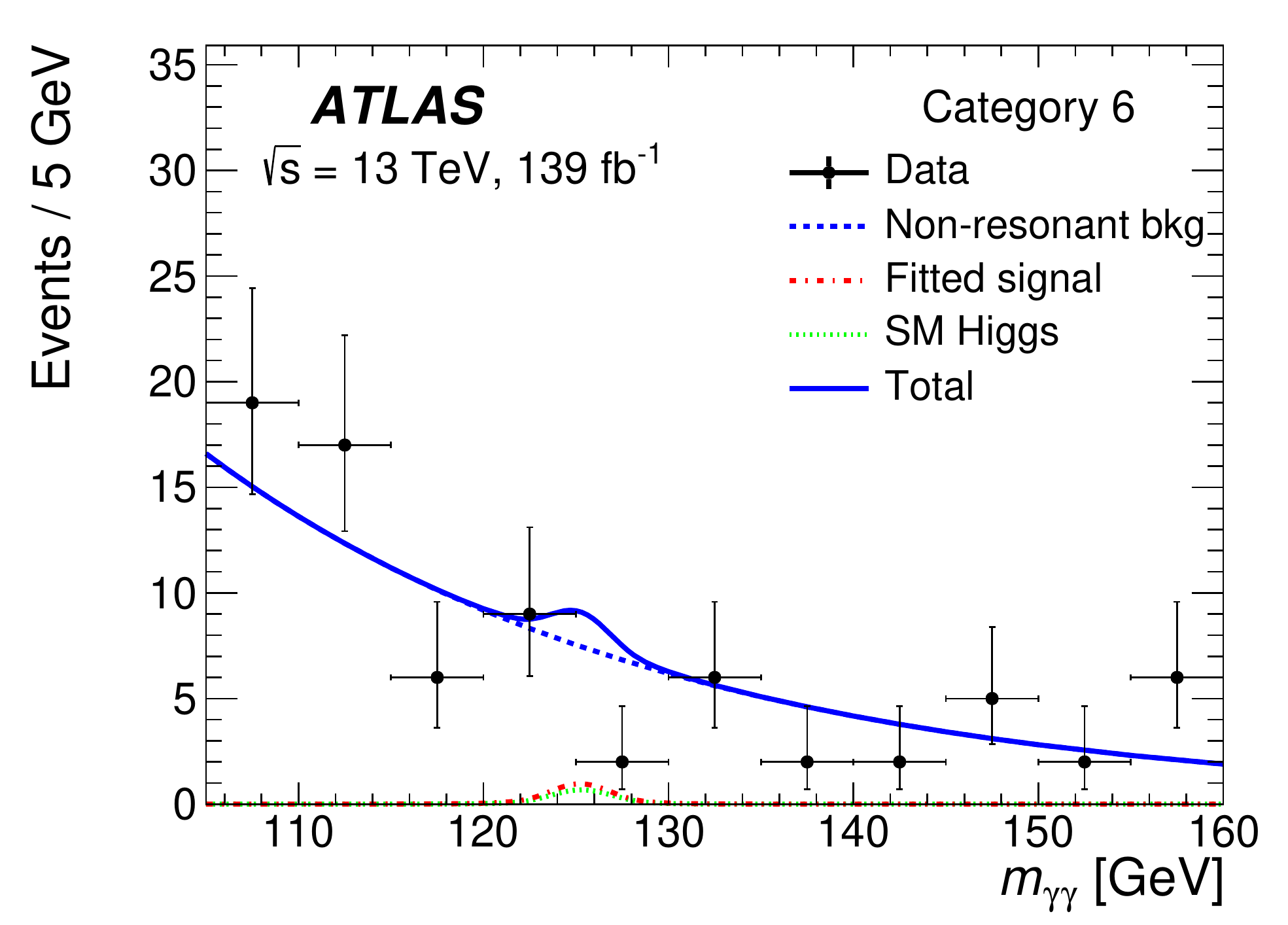}}\\
\subfigure[]{  \includegraphics[width=0.45\textwidth]{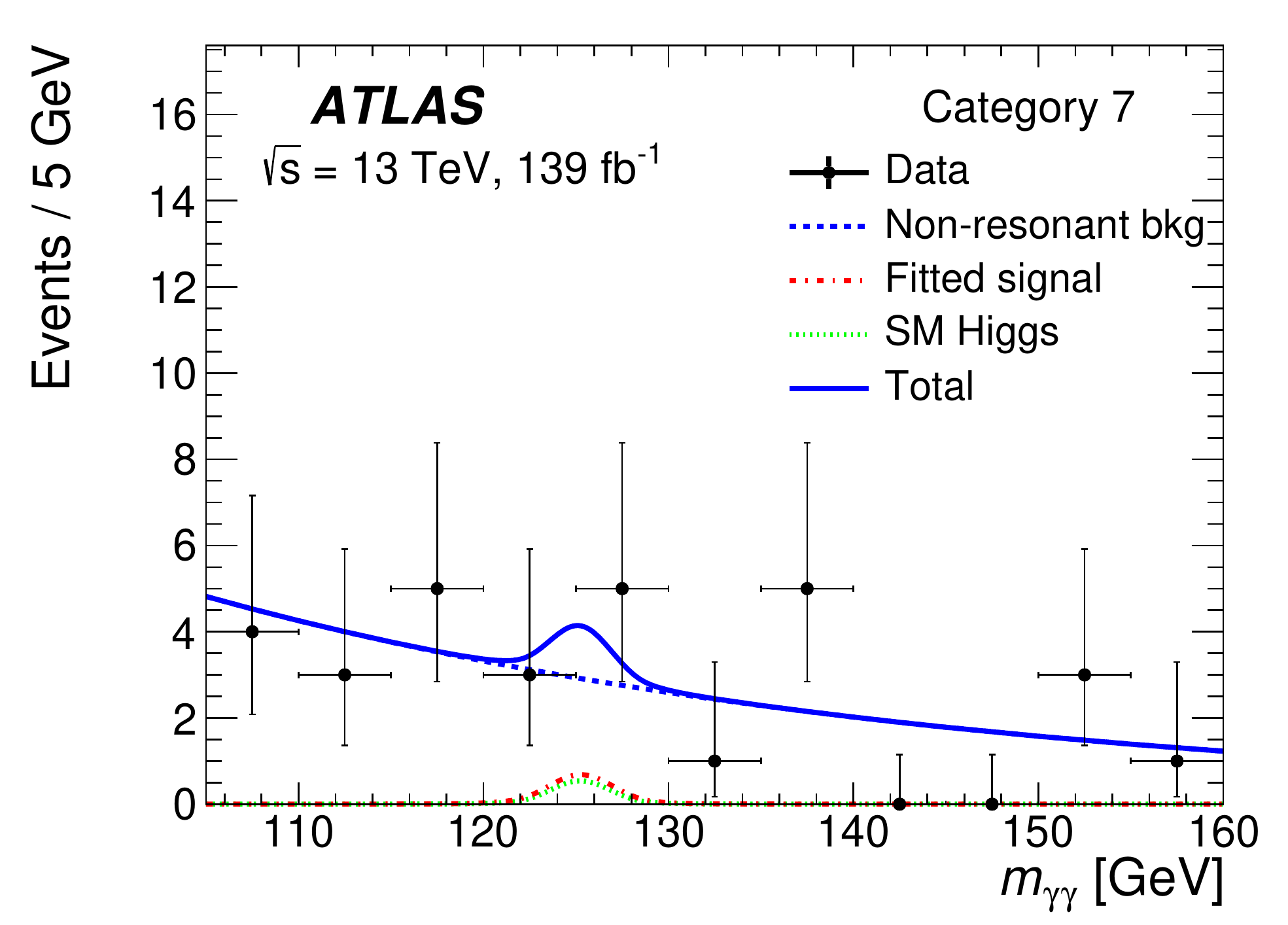}}
\subfigure[]{ \includegraphics[width=0.45\textwidth]{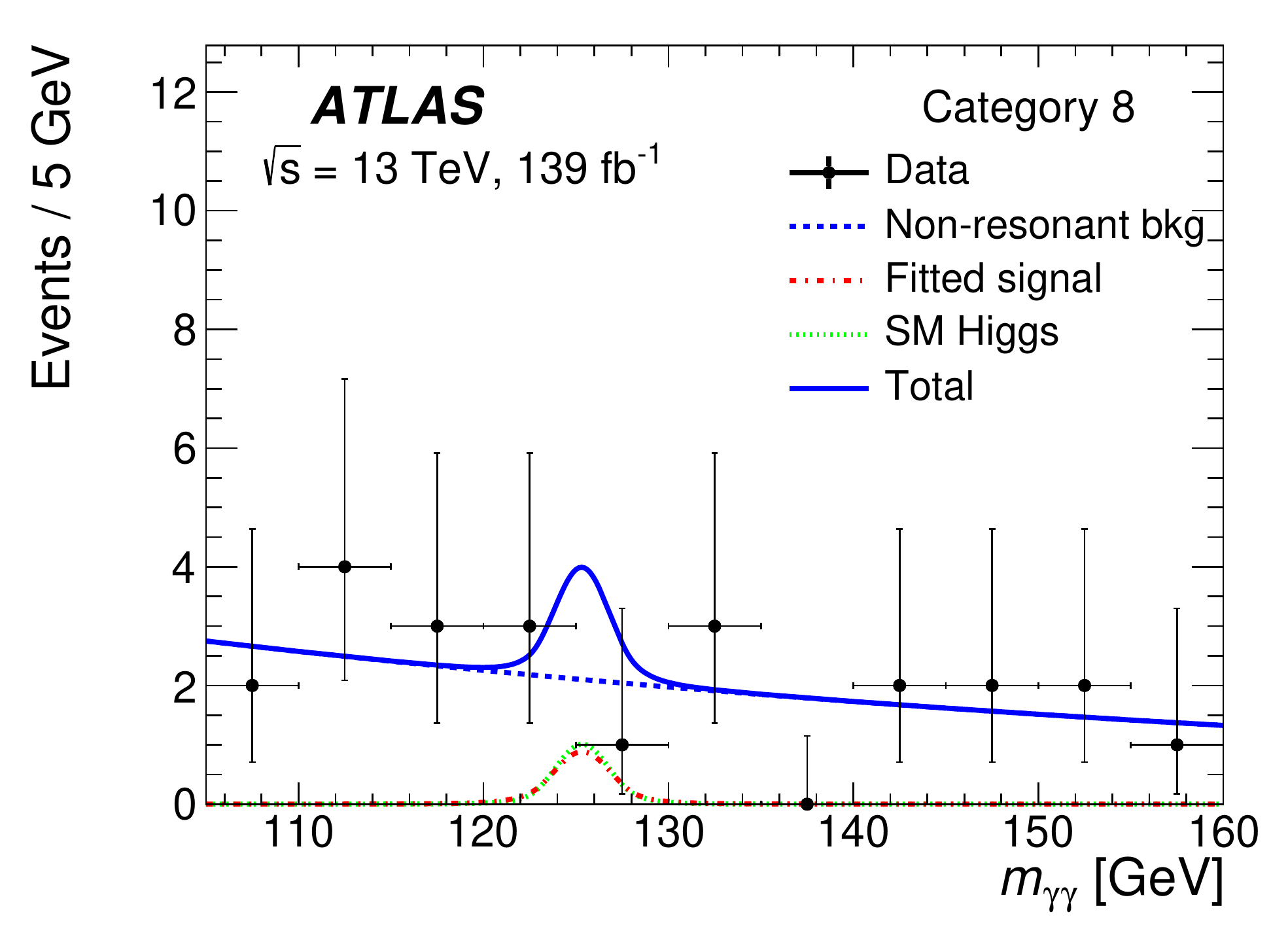}}
\caption{Diphoton invariant mass spectra and the corresponding fitted signal and background in the Hadronic categories (a) 5, (b) 6, (c) 7, and (d) 8.
The signal samples shown correspond to the $\cc\nnn \to W^{\pm}\nn h \nn$ signal with $m(\cc/\nnn)=200~\GeV$ and $m(\nn)=0.5~\GeV$.
The non-resonant background (dashed curve), the SM Higgs boson (dotted curve), and the signal (dash-dotted curve) are obtained
from a simultaneous signal-plus-background fit to the full $m_{\gamma\gamma}$ spectrum for the 12 categories. The total of these contributions is shown by the solid curves.
}
\label{fig:fit_data2}
\end{figure}
 
\begin{figure}[!htbp]
\centering
\subfigure[]{ \includegraphics[width=0.45\textwidth]{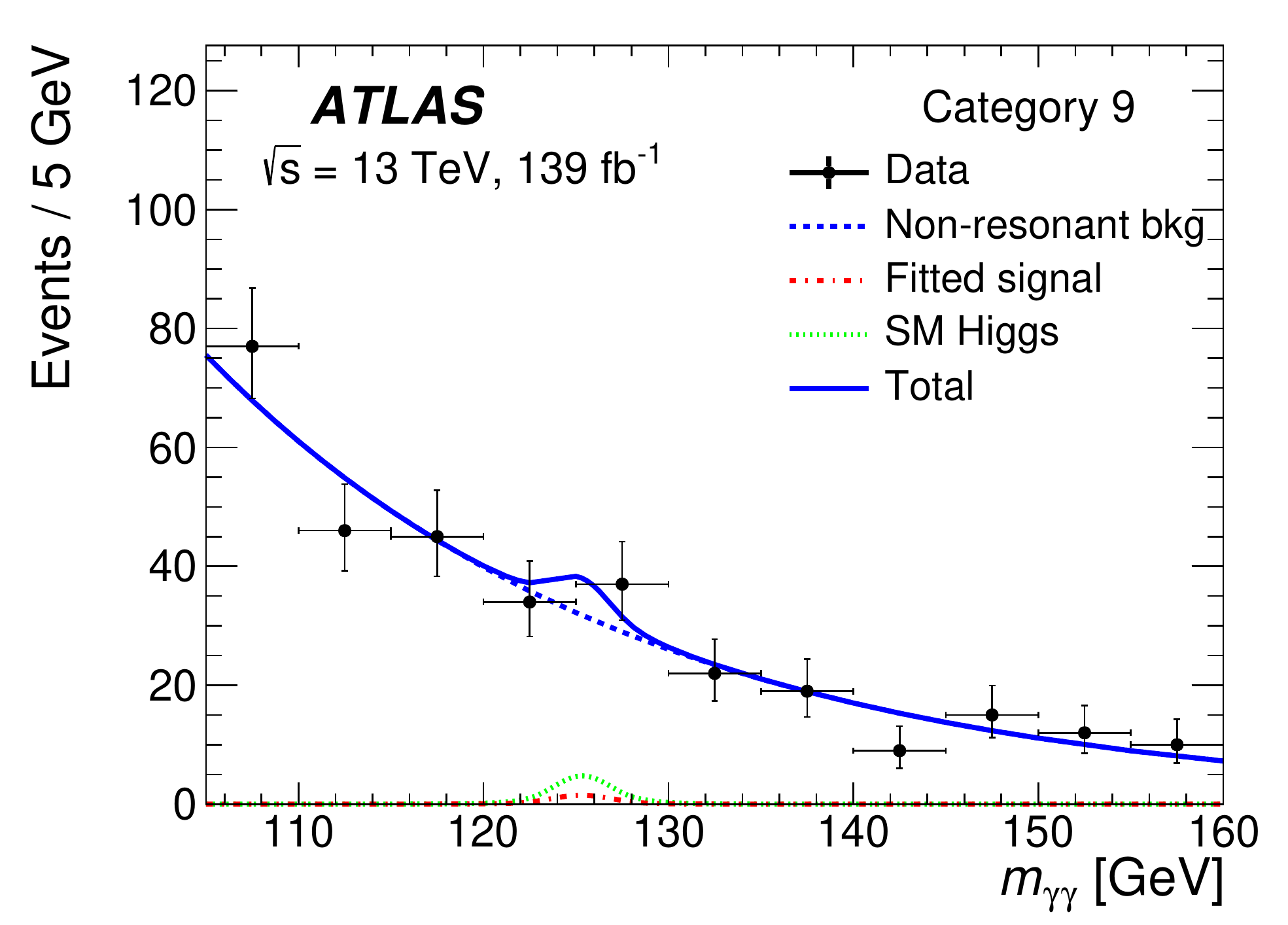}}
\subfigure[]{  \includegraphics[width=0.45\textwidth]{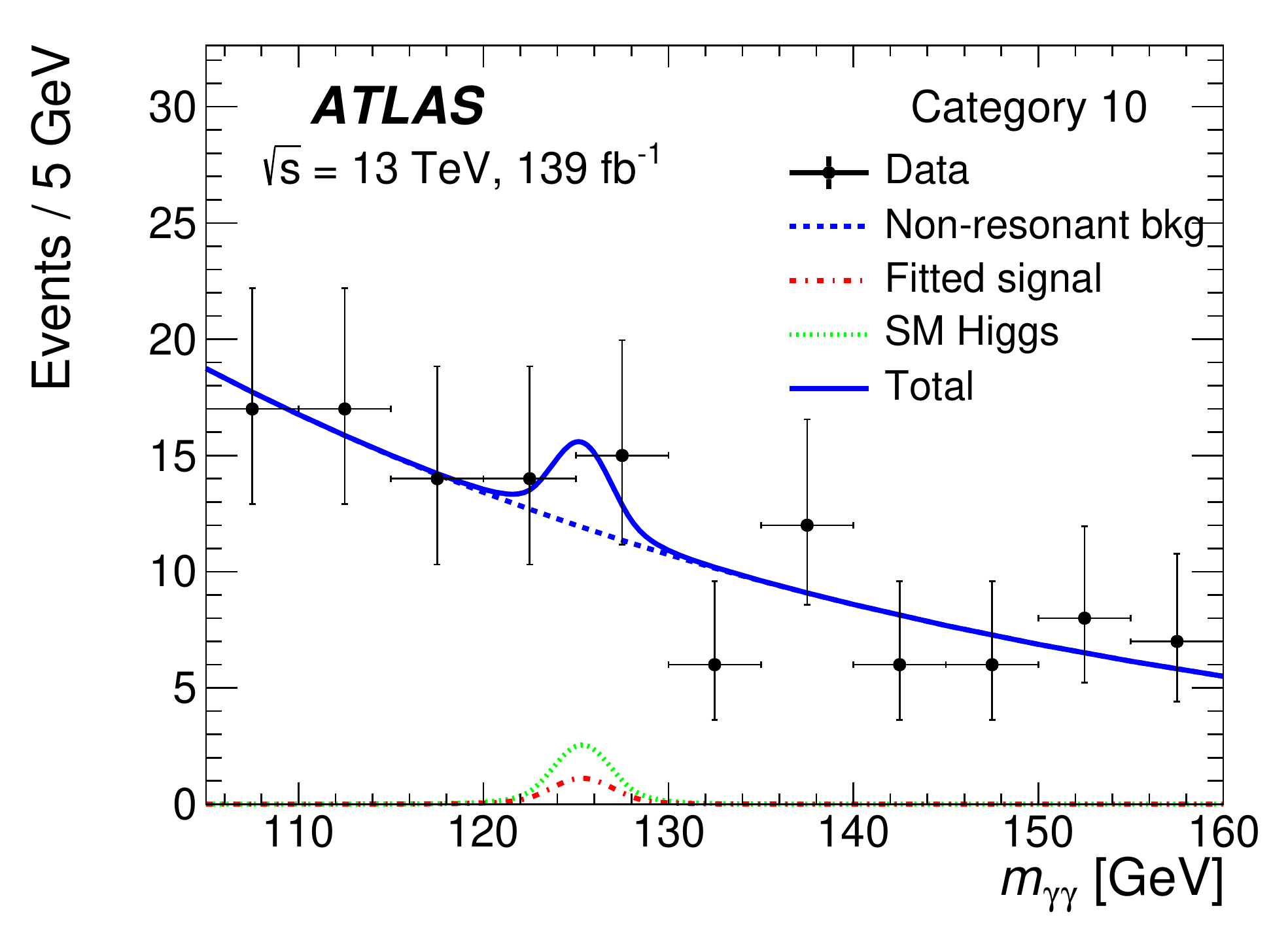}}\\
\subfigure[]{   \includegraphics[width=0.45\textwidth]{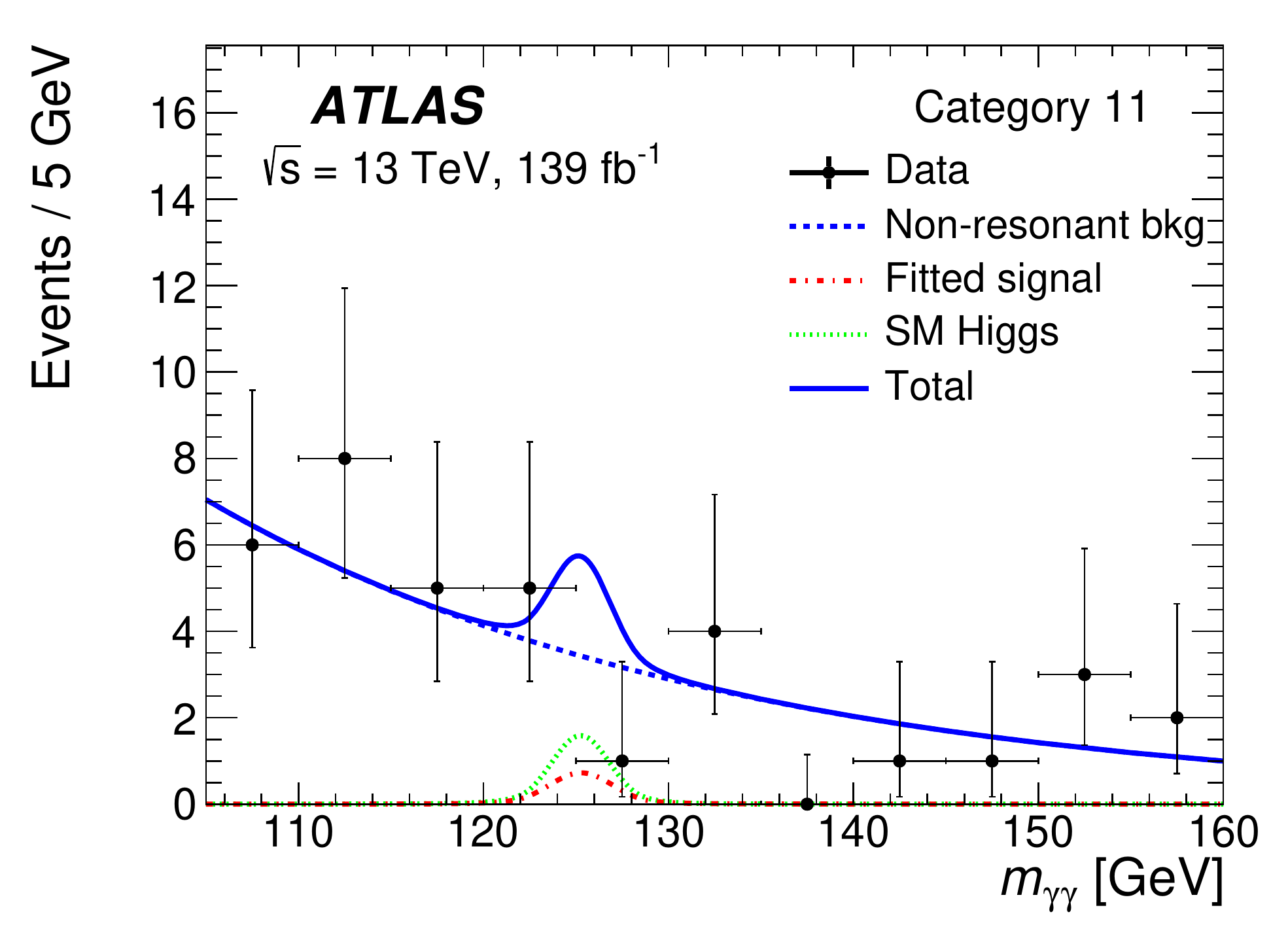}}
\subfigure[]{ \includegraphics[width=0.45\textwidth]{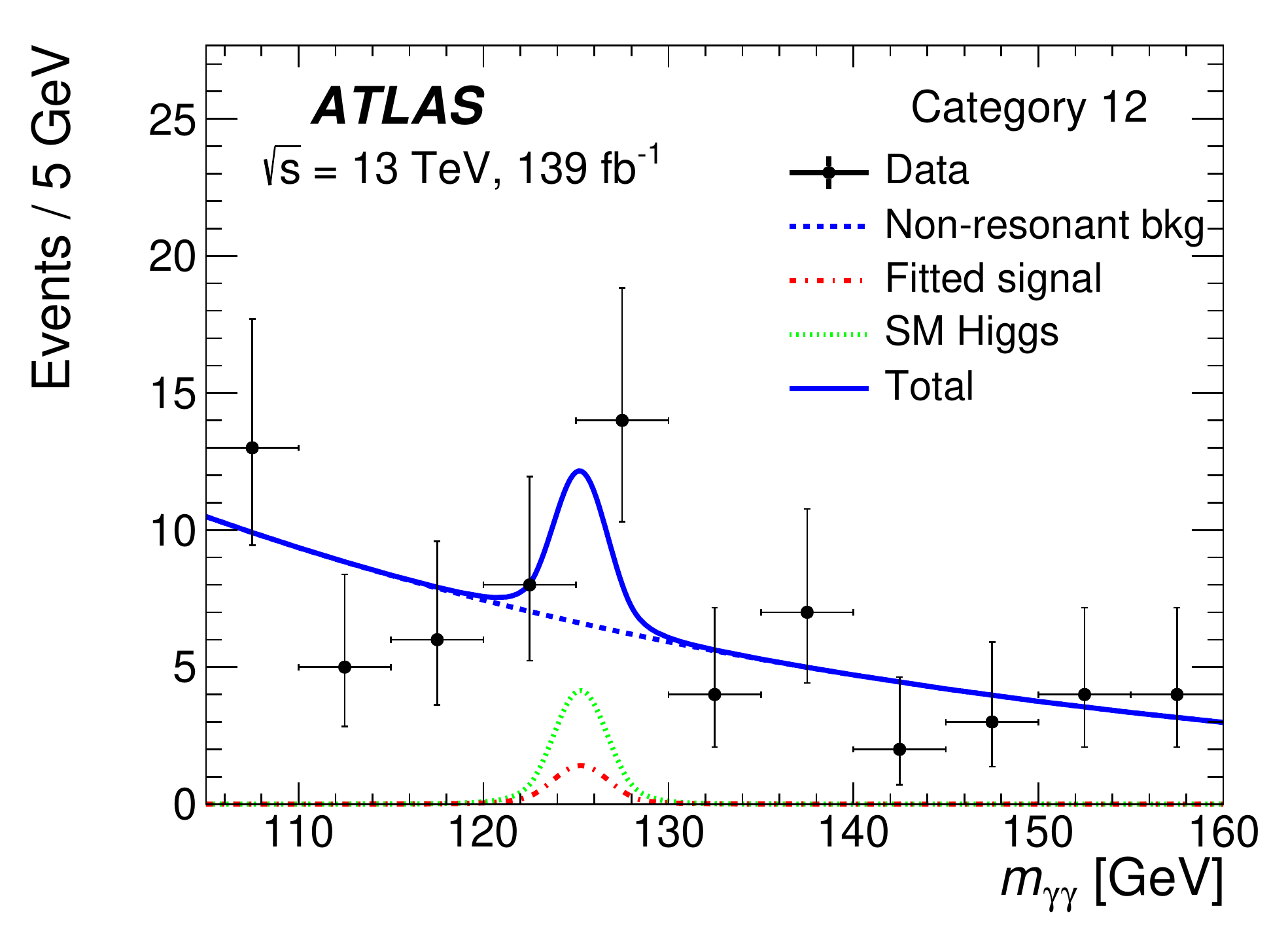}}
\caption{Diphoton invariant mass spectra and the corresponding fitted signal and background in the Rest categories (a) 9, (b) 10, (c) 11, and (d) 12.
The signal samples shown correspond to the $\cc\nnn \to W^{\pm}\nn h \nn$ signal with $m(\cc/\nnn)=200~\GeV$ and $m(\nn)=0.5~\GeV$.
The non-resonant background (dashed curve), the SM Higgs boson (dotted curve), and the signal (dash-dotted curve) are obtained
from a simultaneous signal-plus-background fit to the full $m_{\gamma\gamma}$ spectrum for the 12 categories. The total of these contributions is shown by the solid curves.
}
\label{fig:fit_data3}
\end{figure}
 
Figures~\ref{fig:fit_data1},~\ref{fig:fit_data2} and~\ref{fig:fit_data3} show the \mgg distribution as well as the analytical signal-plus-background fits, for all 12 signal categories.
The total background contains the non-resonant background and the predicted SM Higgs boson contribution.
The fit results combining the $\cc\nnn \to W^{\pm}\nn h \nn$ signal with $m(\cc/\nnn)=200~\GeV$ and $m(\nn)=0.5~\GeV$, SM Higgs boson and non-resonant background are shown as the solid curves. A small excess of around two standard deviations is seen in Category 4, however it is consistent with a statistical fluctuation of the SM prediction.
 
The event yields in the range $120 <\mgg < 130$~\GeV\  for data, the signal models, the SM Higgs boson background and non-resonant background
in the 12 categories are shown in
Table~\ref{tab:final_yields}.
The signal samples shown correspond to the $\cc\nnn \to W^{\pm}\nn h \nn$ signal with $m(\cc/\nnn)=200~\GeV$ and $m(\nn)=0.5~\GeV$,
and the $h\tilde{G} h\tilde{G}$ signal with $m(\nn)=150~\GeV$ and $m(\tilde{G})=1~\MeV$.
The yields for the non-resonant background and the SM Higgs boson are obtained from a simultaneous background-only fit to the full $m_{\gamma\gamma}$ spectrum for the 12 categories.
For the `Leptonic' categories, the $Wh$ process is the largest SM Higgs boson process and occupies 38\%--55\% of total events.
The $t\overline{t}h$ events dominate in the `Hadronic' categories, which account for 36\%--41\% of total SM Higgs boson process events.
In the `Rest' categories, events from the $Zh$ process dominates and holds 37\%--58\% of total SM Higgs boson contribution.
The yields for the signals are estimated from the simulation and normalized to the NLO+NLL predicted cross-sections.
The uncertainties correspond to the statistical and systematic uncertainties summed in quadrature.
For all the categories, data and background predictions agree within the statistical and systematic uncertainties.

\begin{table}[!htbp]
\centering
\caption{Event yields in the range $120 <\mgg < 130$~\GeV\  for
data, the signal models, the SM Higgs boson background and non-resonant background
in each analysis category, for an integrated luminosity of $139$~fb$^{-1}$.
The signal samples shown correspond to the $\cc\nnn \to W^{\pm}\nn h \nn$ signal with $m(\cc/\nnn)=200~\GeV$ and $m(\nn)=0.5~\GeV$,
and the $h\tilde{G} h\tilde{G}$ signals with $m(\nn)=150~\GeV$ and $m(\tilde{G})=1~\MeV$.
The yields for the non-resonant background and SM Higgs boson are obtained from a simultaneous background-only fit to the full $m_{\gamma\gamma}$ spectrum for the 12 categories.
The yields for the signals are estimated from the simulation.
The uncertainties correspond to the statistical and systematic uncertainties summed in quadrature. }
\begin{tabular}{r| r r@{~$\pm$~}l  r@{~$\pm$~}l  r@{~$\pm$~}l r@{~$\pm$~}l r@{~$\pm$~}l r@{~$\pm$~}l r@{~$\pm$~}l}
\hline\hline
Category &  Data      &\multicolumn{2}{c}{Total bkg.} & \multicolumn{2}{c}{Non-resonant bkg.} & \multicolumn{2}{c}{SM Higgs boson} & \multicolumn{2}{c}{$W^{\pm}\nn h \nn$}  &  \multicolumn{2}{c}{$h\tilde{G} h\tilde{G}$}\\
\hline

1 & 258 & 246 & 7 & 230 & 7 & 16.3 & 1.4        & 2.8 & 0.6  & 13 & 6 \\
2 & 85  & 93  & 4& 77 & 4 & 15.6 & 1.3          & 6.6 & 1.5 & 16 & 7 \\
3 & 26  & 24.1  & 2.0 & 17.1 & 1.9 & 7.0 & 0.6  & 6.9 & 1.5  & 6.5 & 2.7 \\
4 & 17  & 12.8  & 1.4 & 8.4 & 1.3 & 4.4 & 0.4   & 10.7 & 2.4  & 3.8 & 1.6 \\
5 & 54  & 60  & 4 & 57.9 & 3.5 & 1.9 & 0.6      & 7.2 & 1.6 & 3.3 & 1.4 \\
6 & 11  & 16.1 & 1.8 & 15.4 & 1.8 & 0.74 & 0.26 & 6.0 & 1.3  & 1.6 & 0.7 \\
7 & 8   & 6.3  & 1.1 & 5.9 & 1.1 & 0.42 & 0.10  & 4.3 & 1.0  & 0.71 & 0.34 \\
8 & 4   & 5.2   & 1.0 & 4.4 & 1.0 & 0.80 & 0.11 & 5.3 & 1.2 & 0.76 & 0.33 \\
9 & 71  & 69  & 4 & 65 & 4 & 3.9 & 0.8          & 9.1 & 2.0  & 3.1 & 1.3 \\
10 & 29 & 26.3  & 2.2 & 24.2 & 2.2 & 2.1 & 0.4  & 6.9 & 1.5 & 1.8 & 0.8 \\
11 & 6  & 8.6   & 1.2 & 7.2 & 1.2 & 1.40 & 0.22 & 4.6 & 1.0  & 1.1 & 0.5 \\
12 & 22 & 16.6 & 1.7 & 13.4 & 1.7 & 3.15 & 0.33 & 7.9 & 1.8 & 1.7 & 0.7 \\
 
\hline\hline
\end{tabular}
\label{tab:final_yields}
\end{table}


The independently fitted $m_{\gamma\gamma}$ distributions for the
`follow-up' signal regions are shown in Figure~\ref{fig:fit_excessFollowUp_SBFit}. No significant excess of events is seen in either of the two regions.
In `SR1L$\gamma\gamma$-a', two events are observed with $3.1 \pm 0.8$ non-resonant background events and $0.5^{+0.2}_{\ \ -0.4}$ SM Higgs boson events expected in the range $120 <m_{\gamma\gamma}<130~\GeV$.
In the case of `SR1L$\gamma\gamma$-b', 31 events are observed, whereas $16.6 \pm 1.9 $ events from non-resonant background and $8.6^{+1.3}_{\ \ -2.1}$ events from the SM Higgs boson are expected in the range $120~<m_{\gamma\gamma}<130~\GeV$.
 
\begin{figure}[!htbp]
\centering
\subfigure[]{\includegraphics[width=0.43\textwidth]{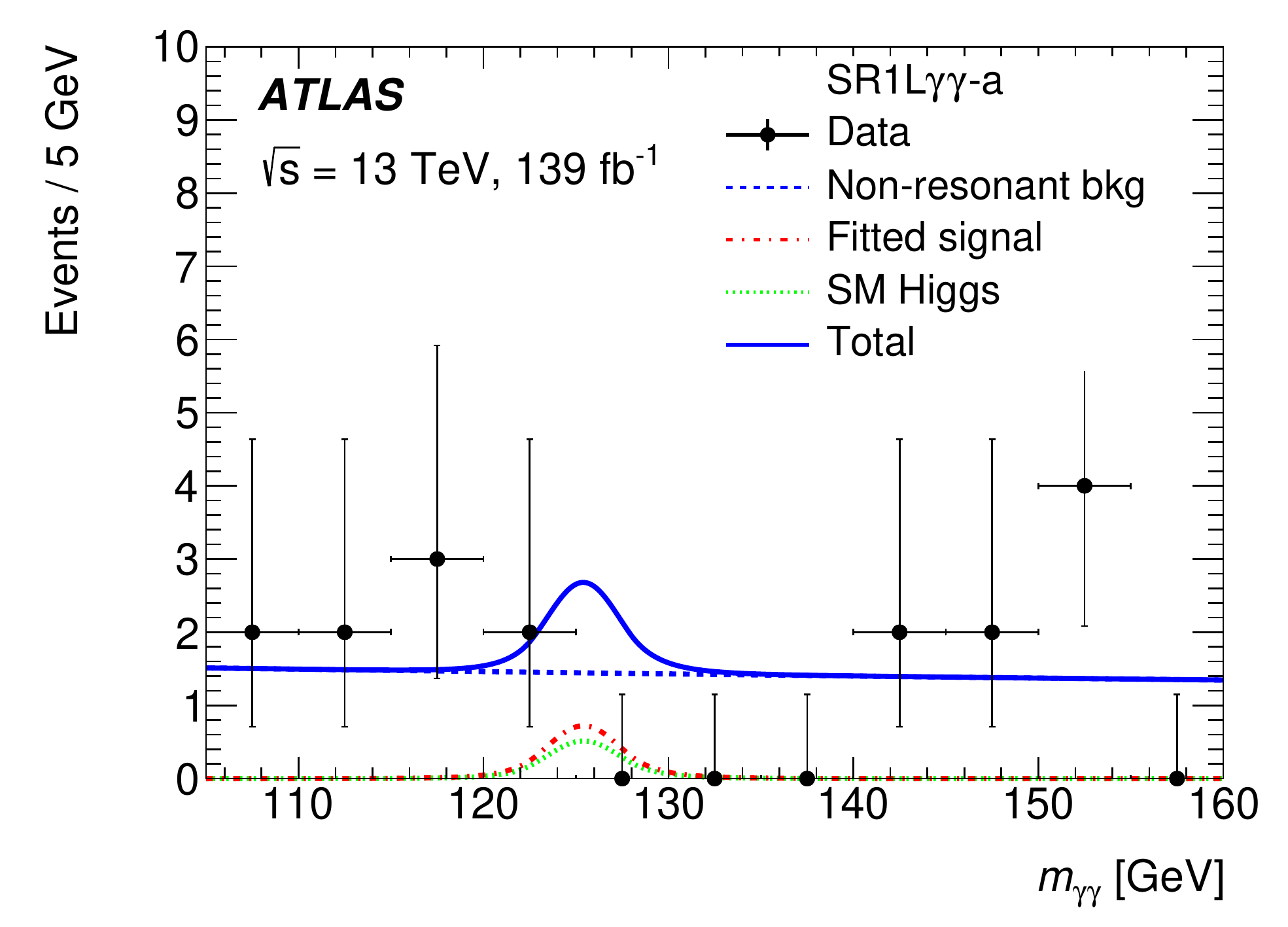}}
\subfigure[]{\includegraphics[width=0.43\textwidth]{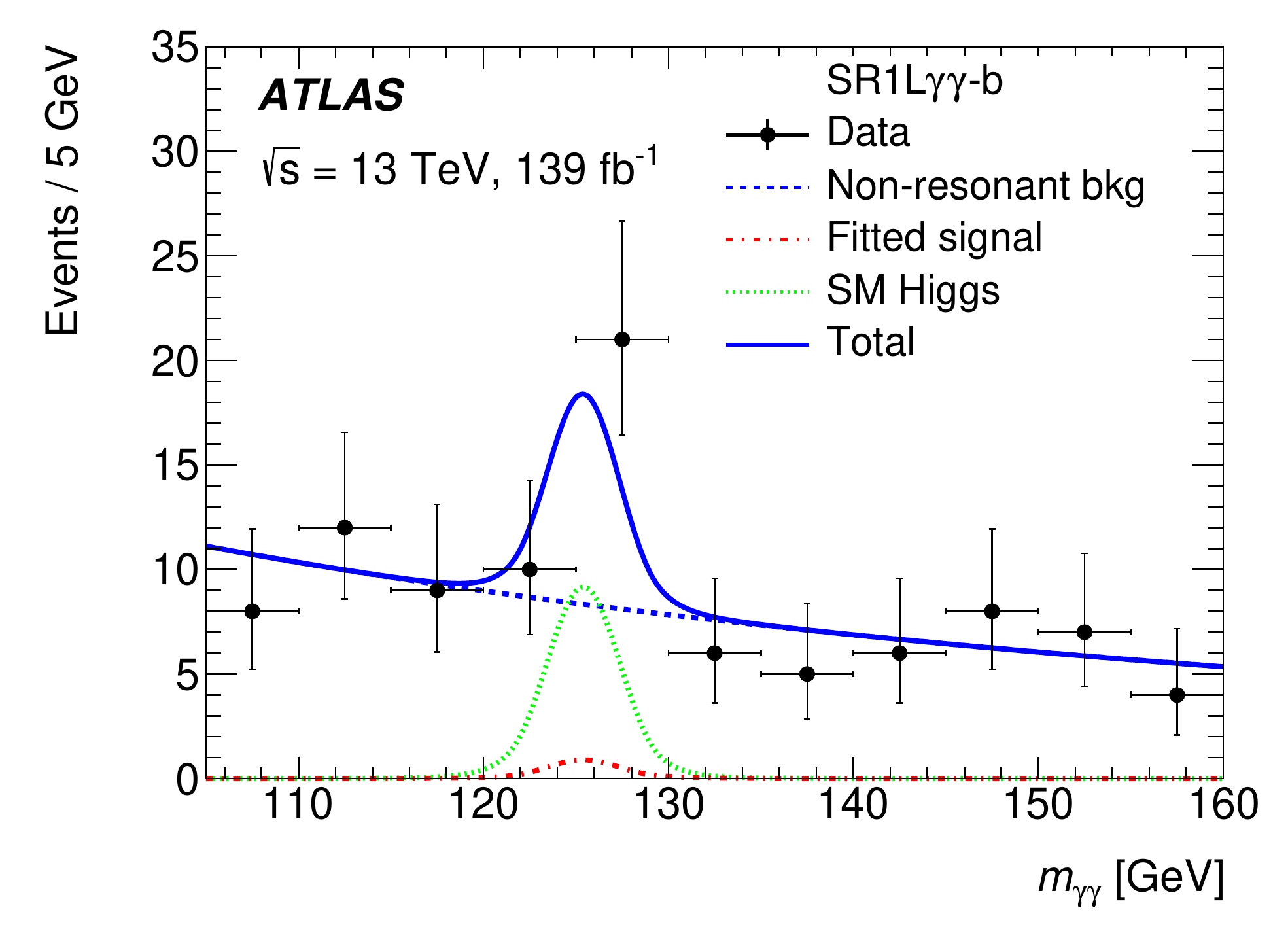}}
\caption{
Diphoton invariant mass spectra and the corresponding fitted signal and background in the signal regions (a) `SR1L$\gamma\gamma$-a' and (b) `SR1L$\gamma\gamma$-b'.
The signal samples shown correspond to the $\cc\nnn \to W^{\pm}\nn h \nn$ signal with $m(\cc/\nnn)=200~\GeV$ and $m(\nn)=0.5~\GeV$.
The non-resonant background (dashed curve), the SM Higgs boson (dotted curve), and the signal (dash-dotted curve) are obtained
from a signal-plus-background fit to the full $m_{\gamma\gamma}$ spectrum in `SR1L$\gamma\gamma$-a' (a) and `SR1L$\gamma\gamma$-b' (b) separately.
The total of these contributions is shown by the solid curves.
}
\label{fig:fit_excessFollowUp_SBFit}
\end{figure}

\subsection{Limits on the visible cross-section}
 
The observed yields agree with the background predictions, as shown in Table~\ref{tab:final_yields},
and no significant excess of events is observed.
Upper limits are set on the visible cross-section $\sigma_{\mathrm{vis}}^{\mathrm{BSM}} \equiv (\cal{A} \times \epsilon \times \sigma)^\mathrm{BSM}$ for BSM physics processes producing $\met$ and an SM
Higgs boson decaying into two photons, where $\cal{A}$ and $\epsilon$ are the acceptance and the efficiency for the signal, respectively.
The limits are extracted by performing a fit to the non-resonant background and SM Higgs boson background, individually for each category, each time injecting a signal with the same mass distribution as the SM Higgs boson but with a free normalisation.
Figure~\ref{fig:model_independent} shows the observed and expected 95\% confidence level (CL) upper
limits on $\sigma_{\mathrm{vis}}^{\mathrm{BSM}}$ for each of the 12 different categories,
which are calculated using a one-sided profile-likelihood ratio and the CL$_\mathrm{s}$
formalism~\cite{Read:2002hq} with the asymptotic approximation described in Ref.~\cite{Cowan:2010js}.
The statistical uncertainty is dominant for all categories.

\begin{figure}[!htbp]
\centering
\includegraphics[width=0.68\textwidth]{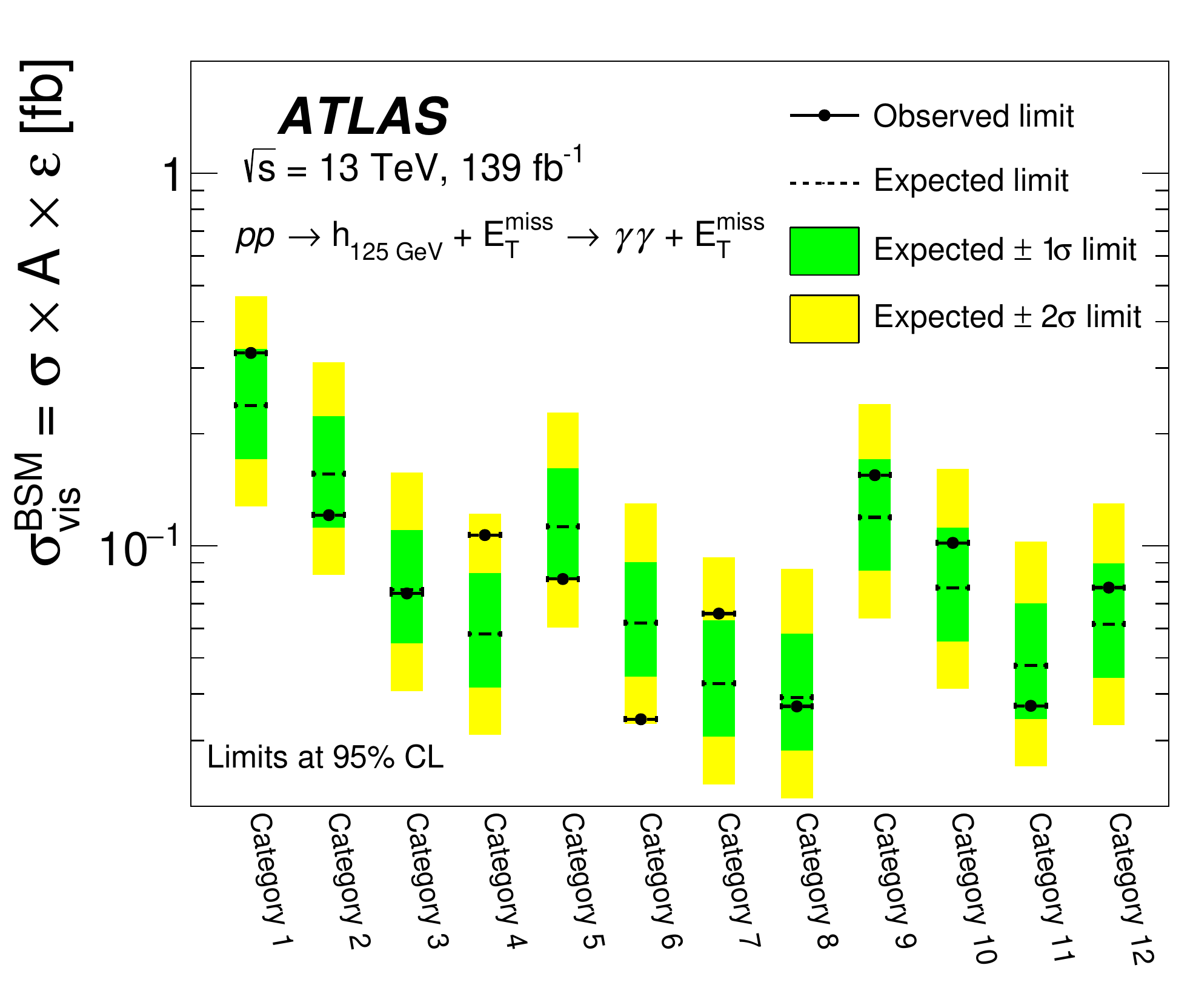}
\caption{The 95\% CL model-independent upper limits computed from individual fits in each of 12 categories on the visible cross-section $\sigma_{\mathrm{vis}}^{\mathrm{BSM}} = \sigma \times A \times \epsilon$ for
any $pp\to h + \met \to \gamma\gamma + \met$  BSM processes.}
\label{fig:model_independent}
\end{figure}

\subsection{Interpretation of the wino-like $\cc\nnn \to  W^{\pm}\nn h \nn$ model }
 
Since no significant excess is observed, fit results are interpreted in terms of 95\% CL exclusion limits on the production cross-section of
the wino-like $\cc\nnn \to  W^{\pm}\nn h \nn$ model~\cite{Alwall:2008ag,Alves:2011wf}.
Upper limits on the contribution of events from the considered processes are computed by using the modified frequentist CL$_\text{s}$ approach based on asymptotic formulae~\cite{Read:2002hq,Cowan:2010js}.
Figure~\ref{fig:xsec_exclusion} shows 95\% CL exclusion limits on the production cross-section of $\cc\nnn \to W^{\pm}\nn h \nn$ as a function of $m(\cc/\nnn)$.
The observed 95\% CL upper limits on the production cross-section vary from
1.92~pb to 0.16~pb for $m(\cc/\nnn)$ from 150~\GeV\ to 600~\GeV. The expected 95\% CL upper limits range from 1.43~pb to 0.11~pb for the same range.
A 95\% CL lower limit of 310~\GeV\ in $m(\cc/\nnn)$, where $m(\tilde{\chi}^{0}_{1})=0.5$~\GeV, is set.
The observed and expected exclusion contours at 95\% CL for the $\cc\nnn$ production in the $m(\cc/\nnn)$--$m(\nn)$ plane are shown in Figure~\ref{fig:mu_2Dexclusion}.
 
\begin{figure}[!htbp]
\centering
\includegraphics[width=0.68\textwidth]{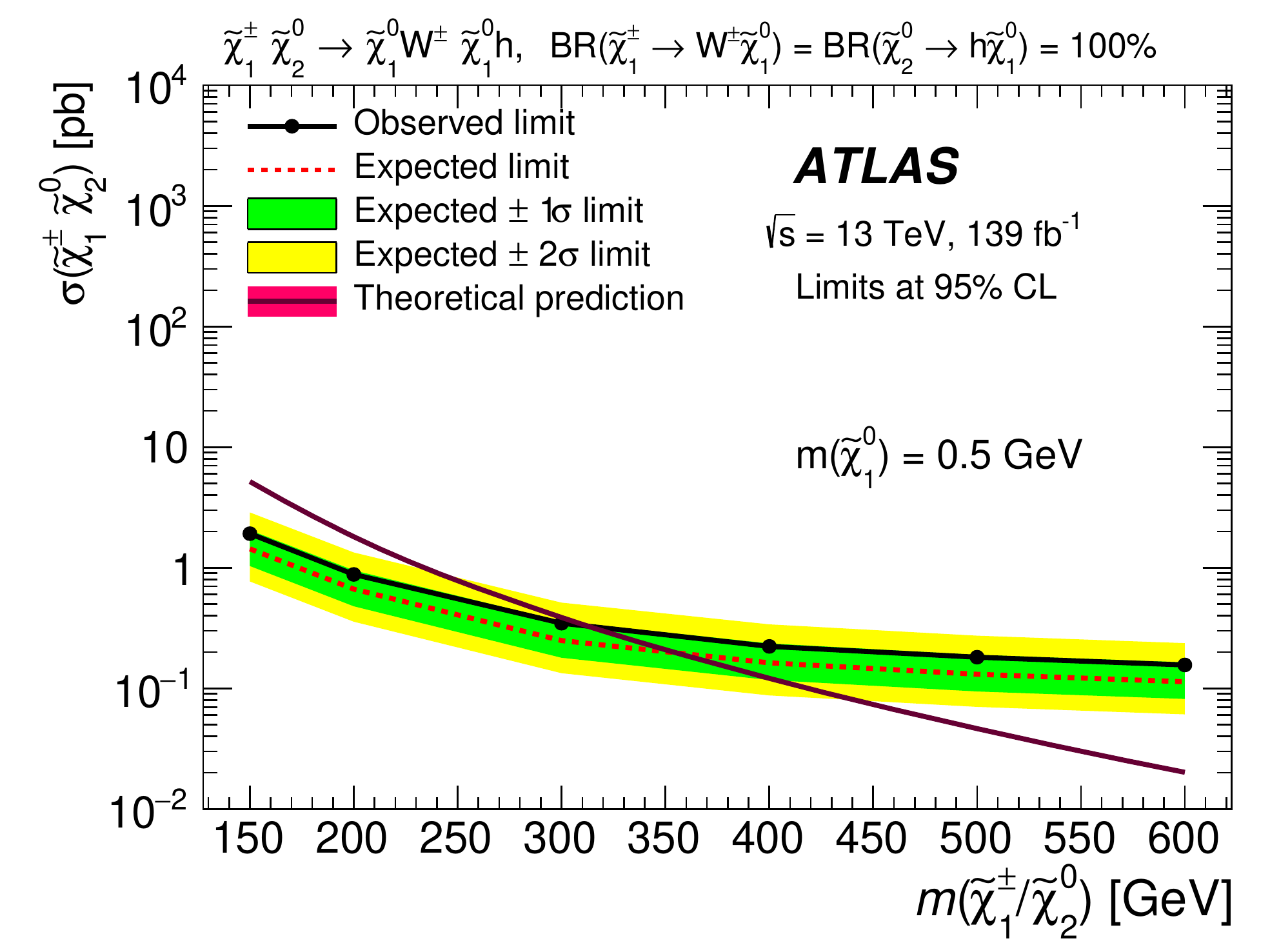}
\caption{Expected and observed $95\%$ CL exclusion upper limits on the production cross-section of $\cc\nnn \to W^{\pm}\nn h \nn$
as a function of $m(\cc/\nnn)$.
}
\label{fig:xsec_exclusion}
\end{figure}

\begin{figure}[!htbp]
\centering
\includegraphics[width=0.68\textwidth]{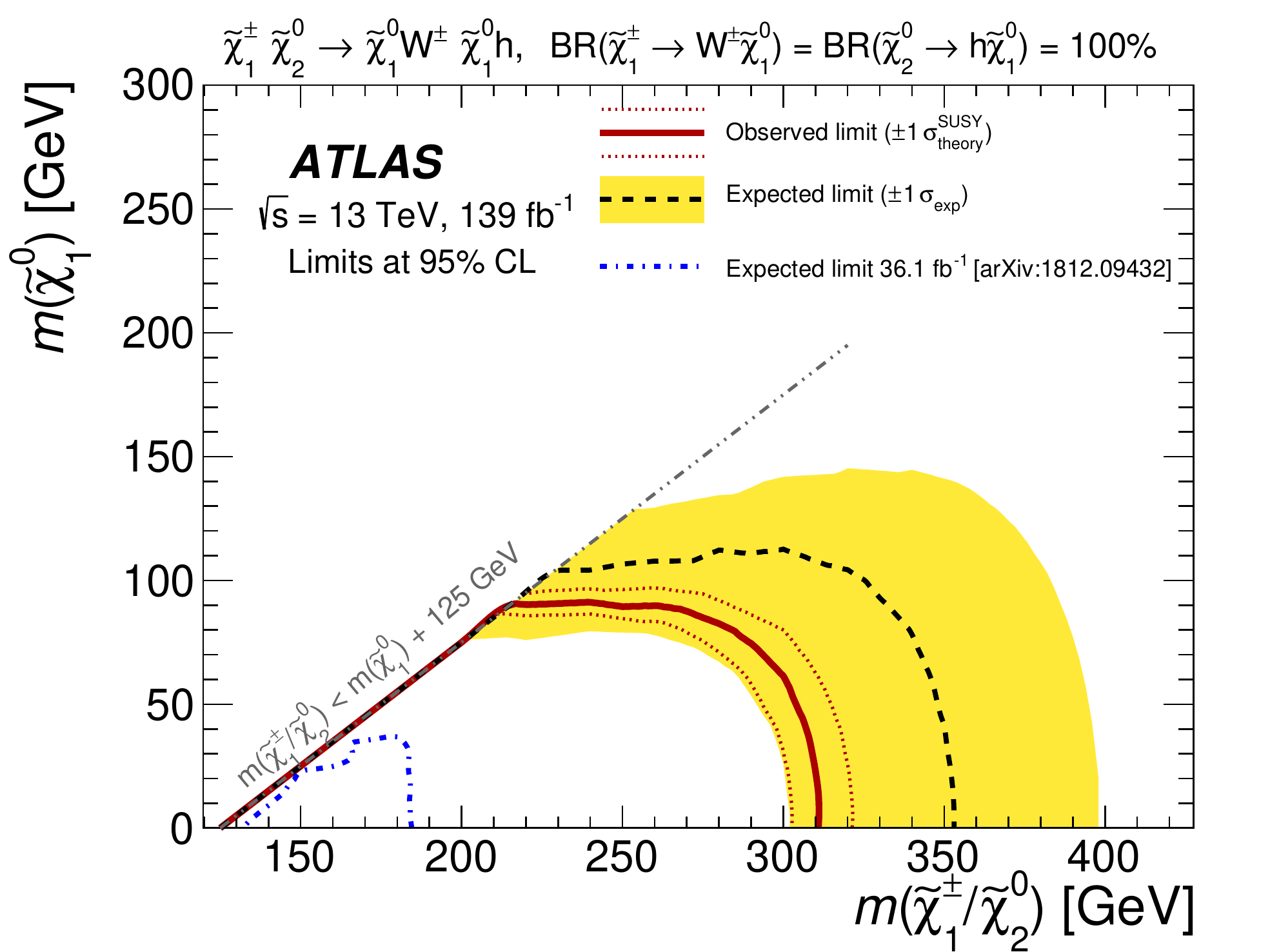}
\caption{
The observed (solid line) and expected (dashed lines) exclusion limit contours at 95\% CL for the $\cc\nnn$ production in the $m(\cc/\nnn)$--$m(\nn)$ plane.
The dotted lines represent the $\pm1\sigma$ theoretical uncertainty for the observed limit. The $\pm1\sigma$ expected exclusion limit contour is shown as the shaded band.
The expected limit for the 36.1~fb$^{-1}$ analysis~\cite{SUSY-2017-01} is also shown for comparison in the dash-dotted line.
}
\label{fig:mu_2Dexclusion}
\end{figure}
 
\subsection{Interpretation of the higgsino-like $h \tilde{G} h \tilde{G}$ model}
 
As a second SUSY scenario, a GMSB model where the two lightest neutralinos and the lightest chargino are higgsinos is considered~\cite{Khachatryan:2014mma,Matchev:1999ft,Ruderman:2011vv}.
The \cc, \nn and \nnn are almost mass degenerate in this model, with \nn being the lightest of the three states. The LSP is a gravitino.
In Figure~\ref{fig:GGM}, the observed and expected 95\% CL upper limits, with uncertainties, on the higgsino production cross-section in the
$h \tilde{G} h \tilde{G}$ models for different $m(\nn)$ masses are presented.
The levelling off of expected limits at low $m(\nn)$ masses is due to the acceptance times efficiency in this region.
The theoretical prediction includes the \nn\nnn, \nn\cc, \nnn\cc, and \cc\ccmp production modes,
where \cc and \nnn promptly decay into the \nn and particles that have too low momentum to be detected.
In the $h \tilde{G} h \tilde{G}$ model, higgsino masses below  380~\GeV\ are excluded at 95\%~CL.
 
\begin{figure}[!htbp]
\centering
\subfigure[]{\includegraphics[width=0.68\textwidth]{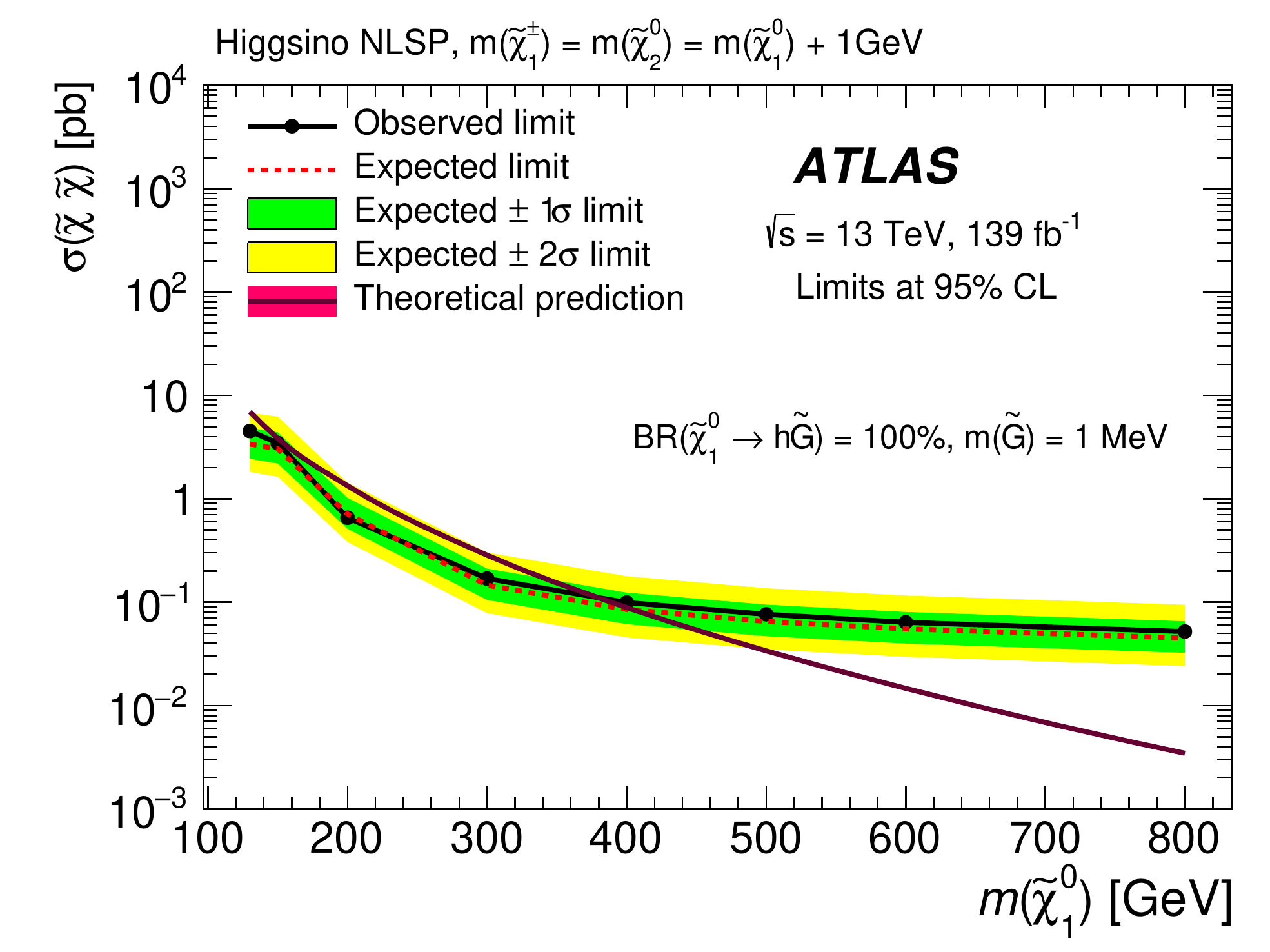}}
\caption{Expected and observed $95\%$ CL exclusion upper limits on the higgsino production ($\tilde{\chi}\tilde{\chi} \equiv \nn\nnn, \nn\cc, \nnn\cc, \cc\ccmp$) cross-section
for the $h \tilde{G} h \tilde{G} $ signal model.
as a function of the higgsino mass.
The theoretical prediction includes the
\nn\nnn, \nn\cc, \nnn\cc, and \cc\ccmp production modes,
where \cc and \nnn promptly decay into the \nn and particles that have too low momentum to be detected.
}
\label{fig:GGM}
\end{figure}

 
\FloatBarrier
 
\clearpage
\section{Conclusion}
\label{sec:conclusion}
 
A search for a chargino and a neutralino decaying via the 125~\GeV\ Higgs boson into photons is presented.
This study is based on the full data collected between 2015 and 2018 with the ATLAS detector at the LHC, corresponding to an integrated luminosity of $139~\mathrm{fb}^{-1}$ of \(pp\) collisions at a centre-of-mass energy of 13~\TeV.
No significant excess over the expected background is observed.
Upper limits at 95\% confidence level are set on the $\cc\nnn$ and higgsino production cross-section, and the visible cross-section for beyond the Standard Model physics processes.
For the $\cc\nnn \to W^{\pm}\nn h \nn$ model, the observed 95\% confidence-level upper limits on the production cross-section vary from
1.92~pb to 0.16~pb for $m(\cc/\nnn)$ from 150~\GeV\ to 600~\GeV, where $m(\tilde{\chi}^{0}_{1})$ is set to  0.5~\GeV.
The expected 95\% confidence-level upper limits range from 1.43~pb to 0.11~pb for the same mass interval.
A 95\% confidence-level lower limit of 310~\GeV\ in $m(\cc/\nnn)$, where $m(\tilde{\chi}^{0}_{1})=0.5$~\GeV, is set.
Upper limits at the 95\% confidence-level are set on the higgsino production cross-section.
Higgsino masses below 380~\GeV\ are excluded for the case of the higgsino fully decaying into a Higgs boson and a gravitino.
 
 
\clearpage
\section*{Acknowledgements}
 

We thank CERN for the very successful operation of the LHC, as well as the
support staff from our institutions without whom ATLAS could not be
operated efficiently.
 
We acknowledge the support of ANPCyT, Argentina; YerPhI, Armenia; ARC, Australia; BMWFW and FWF, Austria; ANAS, Azerbaijan; SSTC, Belarus; CNPq and FAPESP, Brazil; NSERC, NRC and CFI, Canada; CERN; CONICYT, Chile; CAS, MOST and NSFC, China; COLCIENCIAS, Colombia; MSMT CR, MPO CR and VSC CR, Czech Republic; DNRF and DNSRC, Denmark; IN2P3-CNRS and CEA-DRF/IRFU, France; SRNSFG, Georgia; BMBF, HGF and MPG, Germany; GSRT, Greece; RGC and Hong Kong SAR, China; ISF and Benoziyo Center, Israel; INFN, Italy; MEXT and JSPS, Japan; CNRST, Morocco; NWO, Netherlands; RCN, Norway; MNiSW and NCN, Poland; FCT, Portugal; MNE/IFA, Romania; MES of Russia and NRC KI, Russia Federation; JINR; MESTD, Serbia; MSSR, Slovakia; ARRS and MIZ\v{S}, Slovenia; DST/NRF, South Africa; MINECO, Spain; SRC and Wallenberg Foundation, Sweden; SERI, SNSF and Cantons of Bern and Geneva, Switzerland; MOST, Taiwan; TAEK, Turkey; STFC, United Kingdom; DOE and NSF, United States of America. In addition, individual groups and members have received support from BCKDF, CANARIE, Compute Canada and CRC, Canada; ERC, ERDF, Horizon 2020, Marie Sk{\l}odowska-Curie Actions and COST, European Union; Investissements d'Avenir Labex, Investissements d'Avenir Idex and ANR, France; DFG and AvH Foundation, Germany; Herakleitos, Thales and Aristeia programmes co-financed by EU-ESF and the Greek NSRF, Greece; BSF-NSF and GIF, Israel; CERCA Programme Generalitat de Catalunya and PROMETEO Programme Generalitat Valenciana, Spain; G\"{o}ran Gustafssons Stiftelse, Sweden; The Royal Society and Leverhulme Trust, United Kingdom.
 
The crucial computing support from all WLCG partners is acknowledged gratefully, in particular from CERN, the ATLAS Tier-1 facilities at TRIUMF (Canada), NDGF (Denmark, Norway, Sweden), CC-IN2P3 (France), KIT/GridKA (Germany), INFN-CNAF (Italy), NL-T1 (Netherlands), PIC (Spain), ASGC (Taiwan), RAL (UK) and BNL (USA), the Tier-2 facilities worldwide and large non-WLCG resource providers. Major contributors of computing resources are listed in Ref.~\cite{ATL-GEN-PUB-2016-002}.
 

\printbibliography

\clearpage 
 
\begin{flushleft}
\hypersetup{urlcolor=black}
{\Large The ATLAS Collaboration}

\bigskip

\AtlasOrcid[0000-0002-6665-4934]{G.~Aad}$^\textrm{\scriptsize 102}$,    
\AtlasOrcid[0000-0002-5888-2734]{B.~Abbott}$^\textrm{\scriptsize 129}$,    
\AtlasOrcid{D.C.~Abbott}$^\textrm{\scriptsize 103}$,    
\AtlasOrcid[0000-0002-2788-3822]{A.~Abed~Abud}$^\textrm{\scriptsize 71a,71b}$,    
\AtlasOrcid[0000-0002-1002-1652]{K.~Abeling}$^\textrm{\scriptsize 53}$,    
\AtlasOrcid[0000-0002-2987-4006]{D.K.~Abhayasinghe}$^\textrm{\scriptsize 94}$,    
\AtlasOrcid[0000-0002-8496-9294]{S.H.~Abidi}$^\textrm{\scriptsize 167}$,    
\AtlasOrcid[0000-0002-8279-9324]{O.S.~AbouZeid}$^\textrm{\scriptsize 40}$,    
\AtlasOrcid{N.L.~Abraham}$^\textrm{\scriptsize 156}$,    
\AtlasOrcid[0000-0001-5329-6640]{H.~Abramowicz}$^\textrm{\scriptsize 161}$
% ...
\bigskip
\\

% ...
$^{40}$Niels Bohr Institute, University of Copenhagen, Copenhagen; Denmark.\\
% ...
$^{53}$II. Physikalisches Institut, Georg-August-Universit\"{a}t G\"ottingen, G\"ottingen; Germany.\\
% ...
$^{71}$$^{(a)}$INFN Sezione di Pavia;$^{(b)}$Dipartimento di Fisica, Universit\`a di Pavia, Pavia; Italy.\\
% ...
$^{94}$Department of Physics, Royal Holloway University of London, Egham; United Kingdom.\\
% ...
$^{102}$CPPM, Aix-Marseille Universit\'e, CNRS/IN2P3, Marseille; France.\\
$^{103}$Department of Physics, University of Massachusetts, Amherst MA; United States of America.\\
% ...
$^{129}$Homer L. Dodge Department of Physics and Astronomy, University of Oklahoma, Norman OK; United States of America.\\
% ...
$^{156}$Department of Physics and Astronomy, University of Sussex, Brighton; United Kingdom.\\
% ...
$^{161}$Raymond and Beverly Sackler School of Physics and Astronomy, Tel Aviv University, Tel Aviv; Israel.\\
% ...
$^{167}$Department of Physics, University of Toronto, Toronto ON; Canada.\\
% ...
$^{a}$ Also at Borough of Manhattan Community College, City University of New York, New York NY; United States of America.\\
$^{b}$ Also at CERN, Geneva; Switzerland.\\
% ...

\end{flushleft}


\end{document}